\documentclass[%
 reprint,
 amsmath,amssymb,
 aps,prl,superscriptaddress
]{revtex4-2}

\usepackage{graphicx}
\usepackage{dcolumn}
\usepackage{bm}
\usepackage{xcolor}
\usepackage{physics}
\usepackage{ragged2e}
\usepackage{bbm}
\usepackage[colorlinks=true, allcolors=blue]{hyperref}
\usepackage{subfigure}

\begin{document}
\newcommand{\comment}[1]{{\color{red}#1}}
\newcommand{\todo}[1]{{\color{black}#1}}
\newcommand{\topic}[1]{\newline\noindent\textbf{#1}:}

\preprint{APS/123-QED}

\title{Anomalous Neutron Nuclear-Magnetic Interference Spectroscopy}
\affiliation{%
 Quantum Measurement Group, MIT, Cambridge, MA 02139, USA}
\affiliation{Department of Nuclear Science and Engineering, MIT, Cambridge, MA 02139, USA}
\affiliation{Department of Physics, MIT, Cambridge, MA 02139, USA}

\affiliation{Mechanical Engineering and Materials Science and Engineering, University of California, Riverside, Riverside, CA, 92521, USA}
\affiliation{Department of Chemistry, Michigan State University, East Lansing, MI 48824, USA}

\author{Chuliang Fu}
\affiliation{%
 Quantum Measurement Group, MIT, Cambridge, MA 02139, USA}
\affiliation{Department of Nuclear Science and Engineering, MIT, Cambridge, MA 02139, USA}
\affiliation{These authors contributed equally to this work.}

\author{Phum Siriviboon}%
\affiliation{%
 Quantum Measurement Group, MIT, Cambridge, MA 02139, USA}
\affiliation{Department of Physics, MIT, Cambridge, MA 02139, USA}
\affiliation{These authors contributed equally to this work.}

\author{Artittaya Boonkird}%
\affiliation{%
 Quantum Measurement Group, MIT, Cambridge, MA 02139, USA}
\affiliation{Department of Nuclear Science and Engineering, MIT, Cambridge, MA 02139, USA}
\affiliation{These authors contributed equally to this work.}

\author{Michael Landry}
\affiliation{%
 Quantum Measurement Group, MIT, Cambridge, MA 02139, USA}
\affiliation{Department of Nuclear Science and Engineering, MIT, Cambridge, MA 02139, USA}
\affiliation{Department of Physics, MIT, Cambridge, MA 02139, USA}

\author{Chen Li}
\affiliation{Mechanical Engineering and Materials Science and Engineering, University of California, Riverside, Riverside, CA, 92521, USA}
\author{Weiwei Xie}
\affiliation{Department of Chemistry, Michigan State University, East Lansing, MI 48824, USA}
\author{Mingda Li}%
\thanks{Corresponding authors. \href{mailto:mingda@mit.edu}{mingda@mit.edu}}
\affiliation{%
 Quantum Measurement Group, MIT, Cambridge, MA 02139, USA}
\affiliation{Department of Nuclear Science and Engineering, MIT, Cambridge, MA 02139, USA}

\date{\today}

\begin{abstract}
The electron-phonon interaction plays a critical role in materials’ electrical, thermal, optical, and superconducting properties. However, measuring the phonon mode-resolved electron-phonon interaction has been challenging. Here we propose neutron-scattering-based Anomalous Neutron nUclear-Magnetic Interference Spectroscopy (ANUBIS), where the co-existence of neutron nuclear scattering and magnetic scattering leads to anomalous dynamical structure factor under the presence of the electron-phonon interaction. Such anomalous structure factor is linear in electron-phonon coupling constant at the phonon wavevector, and is directly proportional to the momentum and energy-resolved dielectric function. The experimental configuration can be achieved using existing polarized inelastic neutron scattering setup, and an order-of-magnitude estimate shows the viability to observe the anomalous scattering signal is around $10^{-4}$ to $10^{-3}$ relative to phonon scattering, which is achievable at emerging neutron facilities. Our proposal offers an alternative neutron-based metrology to probe the crucial electronic properties. 
\end{abstract}

\maketitle


Neutron scattering is a powerful and versatile technique to probe the atomic and magnetic structures and dynamics in materials. Since its inception, neutron scattering has been characterized by two primary types: nuclear scattering and magnetic scattering \cite{lovesey1984}. This distinction arises because the interaction between thermal neutrons and matter is primarily governed by two mechanisms: nuclear interaction with atomic nuclei and magnetic interaction with electron spins. This naturally leads to the bi-partite of the nuclear dynamical structure factor and the magnetic dynamical structure factor as two fundamental measurables. Despite the awareness of their interactions in polarized-neutron magnetic scattering \cite{SergeiVMaleev_2002}, the implication of the interference between nuclear scattering and magnetic scattering is well understood beyond the magnetoelastic interactions \cite{aliouane2006field,avci2011magnetoelastic}.  

Over past decades, neutron scattering has demonstrated remarkable efficacy in exploring a diverse range of materials. This includes energy materials like hydrogen storage materials \cite{ramirez2009neutron} and ionic conductors \cite{niedziela2019selective}, ceramics \cite{sun2022mutual,sun2023spin}, soft matters such as polymers \cite{chu2001small} and biological macromolecules \cite{ashkar2018neutron}, quantum materials such as superconductors \cite{shirane1969lattice,tranquada2020cuprate,mcqueeney1999anomalous,ma2017prominent,yildirim2001giant}, quantum spin ice and quantum spin liquids \cite{ross2011quantum,stone2006quasiparticle,banerjee2018excitations,han2012fractionalized}, as well as topological materials \cite{gaudet2021weyl,drucker2023topology,bhattacharjee2022interface,nguyen2020topological}. However, the measurement of the electron-phonon interaction and dielectric functions has posed a significant challenge for neutron scattering and other techniques in general. For one thing, the electron-phonon interaction deduced from existing neutron scattering techniques is largely mode-averaged \cite{yildirim2001giant}, while emerging techniques such as inelastic spin-echo measure the phonon linewidth have an extremely high technical barrier and quadratic dependence on the electron-phonon interaction \cite{keller2006momentum,li2014novel}, requiring orders-of-magnitude more sample quantity. For the other thing, angular-resolved photoemission spectroscopy (ARPES) can measure the electron-phonon interaction, but only at the electron wavevector \cite{de2020direct}, while electron energy loss spectroscopy (EELS) measures the dielectric function at finite frequency but zero momentum \cite{egerton2011electron}. Given the significance of the electron-phonon interaction and the dielectric function to materials properties, the development of an alternative and simpler method for their measurement would be highly desirable.

In this \emph{Letter}, we theoretically propose a novel experimental scheme to measure the mode-resolved electron-phonon coupling constant and full momentum- and energy-resolved dielectric function. With the incidence of polarized neutrons and the presence of electron-phonon interaction, the neutron nuclear interaction and the magnetic interaction generate a previously overlooked interference term, resulting in an anomalous scattering structure factor, which we term as Anomalous Neutron nUclear-Magnetic Interference Spectroscopy (ANUBIS). The signal strength of this anomalous scattering structure factor with the linear dependence of the magnetic field is estimated with around $10^{-4}$ to $10^{-3}$ order-of-magnitude with respect to phonon scattering, which is within the feasibility of observation by emerging neutron facilities such as Second Target Station (STS), European Spallation Source (ESS), or upgraded Spallation Neutron Source (SNS). Since neutron scattering is seldom used to directly probe electronic properties except for rare cases \cite{koitzsch2016nesting}, our theoretical proposal can broaden the working scope of neutron scattering.

\textit{Anomalous dynamical structure factor.} \textemdash 
We begin by showing the origin of the anomalous dynamical structure factor within the crystalline materials. The neutron-matter interaction consists of both neutron nuclear interaction and neutron-magnetic interaction, and can be written in a neutron spinor form as: 
\begin{equation}\label{eq:int-Hamiltonian}
H_I=\frac{g}{L^3} \sum_{\mathbf{k} \mathbf{q}} \Psi_{N \mathbf{k}+\mathbf{q}}^{\dagger}\left(\rho_{\mathbf{q}}-\gamma r_e \boldsymbol{\sigma}_N \cdot \mathbf{M}_{\perp \mathbf{q}}\right) \Psi_{N \mathbf{k}}
\end{equation}
where $g=\frac{2 \pi \hbar^2}{m_N}$ is the nuclear interaction strength, $\gamma=+1.913$ is the neutron dimensionless gyromagnetic ratio, $r_e$ is the classical radius of electrons, $\Psi_{N \vb{k}}=\left[\Psi_{N \vb{k}\uparrow}, \Psi_{N \vb{k} \downarrow}\right]^T$ is the neutron field operator satisfying $\left\{\Psi_{N \vb{k} \sigma}, \Psi_{N \vb{k}^{\prime} \sigma^{\prime}}^{\dagger}\right\}=\delta_{\vb{k} \vb{k}^{\prime}} \delta_{\sigma \sigma^{\prime}}$. The advantage of this spinor formulation is its effective handling of both spin non-flip and spin-flip scattering, as well as its applicability to both polarized and non-polarized neutrons. For nuclear interaction, we have $\rho_{\mathbf{q}}=\left(\begin{array}{cc}\rho_{\mathbf{q} \uparrow} & 0 \\ 0 & \rho_{\mathbf{q} \downarrow}\end{array}\right), \quad$ and $\rho_{\mathbf{q} \uparrow / \downarrow}=\sum_i b_{i \uparrow / \downarrow} e^{-i \mathbf{q} \cdot \mathbf{R}_i} \quad$ is the scattering-length-density-weighted nuclear density for spin up $(\uparrow)$ and spin down $(\downarrow)$ neutrons, respectively. When the nuclear spin effect of scattering length density is neglected, we have $\rho_{\mathbf{q} \uparrow}=\rho_{\mathbf{q} \downarrow}$.
For magnetic scattering, $\boldsymbol{\sigma}_N$ is the neutron Pauli matrix, and in this work we focus on itinerant spins with $\mathbf{M}_{\perp \mathbf{q}}=\sum_{\mathbf{k}^{\prime}} c_{\mathbf{k}^{\prime}}^{\dagger} \mathbf{m}_{\perp}(\mathbf{q}) c_{\mathbf{k}^{\prime}+\mathbf{q}}$, where $\mathbf{m}_{\perp}(\mathbf{q})=\frac{\mathbf{q} \times \mathbf{s} \times \mathbf{q}}{q^2}-i \mathbf{k}_e I_{2 \times 2} \times \frac{\mathbf{q}}{q^2}$ and $c_{\mathbf{k}}=\left(c_{\mathbf{k} \uparrow}, c_{\mathbf{k} \downarrow}\right)^T$ is the electron spinor, though the work is equally applicable to localized spins where $\mathbf{M}_{\perp \mathbf{q}}=\sum_i e^{i \mathbf{q} \cdot \mathbf{R}_i} \frac{\mathbf{q} \times \mathbf{s}_i \times \mathbf{q}}{q^2}$. From Eq.\,\eqref{eq:int-Hamiltonian}, it can be shown that the total dynamical structure factor can be written as (see Supplemental Information A \cite{supp})
\begin{align}
S_{\text {tot }}(\mathbf{q}, E)= & S_{\rho \rho}(\mathbf{q}, E)+\left(\gamma r_e\right)^2 S_{M M}(\mathbf{q}, E) 
\nonumber \\
& -2 \gamma r_e S_{\text {ano }}(\mathbf{q}, E)
\end{align}
in which we have the nuclear dynamical structure factor
\begin{equation}
\begin{split}
S_{\rho \rho}(\mathbf{q}, E)& =\left(\begin{array}{cc}
S_{\rho \rho, \uparrow \uparrow}(\mathbf{q}, E) & 0 \\
0 & S_{\rho \rho, \downarrow \downarrow}(\mathbf{q}, E)
\end{array}\right) \\
 S_{\rho \rho, \uparrow \uparrow}(\mathbf{q}, E)&=\int_{-\infty}^{+\infty}\left\langle\rho_{\mathbf{q} \uparrow}(t) \rho_{-\mathbf{q} \uparrow}(0)\right\rangle_H e^{i E t / \hbar} d t
 \label{eq:S-rho}
\end{split}
\end{equation}
and magnetic dynamical structure factor
\begin{align}
\label{eq:S-mag}
    S_{M M}(\mathbf{q}, E)& = \begin{pmatrix}
    S_{M^z M^z}(\bm{q}, E) & S_{M^- M^+}(\bm{q}, E)\\
    S_{M^+ M^-}(\bm{q}, E) & S_{M^z M^z}(\bm{q}, E)
    \end{pmatrix} \nonumber 
\\
 S_{M^j M^l}(\mathbf{q}, E)& =\int_{-\infty}^{\infty}\left\langle M_{\mathbf{q} \perp}^j(t) M_{-\mathbf{q} \perp}^l(0)\right\rangle_H e^{i E t / \hbar} d t  \nonumber 
 \\
 M^{\pm} &= M^x \mp i M^y
 \\
& (j,l=+, -, z) \nonumber 
\end{align}
where this formalism is fully reducible to the common double differential cross section (see Supplemental Information B \cite{supp}). 

The anomalous dynamical structure factor can be written as
\begin{equation}
\begin{split}
S_{\text {ano }}(\mathbf{q}, E) &= 
 \left(\begin{array}{cc}
S_{\rho M \uparrow}^z(\mathbf{q}, E) & 0 \\
0 & -S_{\rho M \downarrow}^z(\mathbf{q}, E)
\end{array}\right) \\
 S_{\rho M \uparrow / \downarrow}^z(\mathbf{q}, E)&=\int_{-\infty}^{+\infty}\left\langle\rho_{\mathbf{q} \uparrow / \downarrow}(t) M_{-\mathbf{q} \perp}^z(0)\right\rangle_H e^{i E t / \hbar} d t
 \label{eq:S-ano}
\end{split}
\end{equation}
which originates from the interference between nuclear and magnetic interactions. The diagonal and off-diagonal part in $S_{\text {ano }}(\mathbf{q}, E)$ represents the spin-non-flip scatterings and spinflip scattering, respectively. For spin-non-flip scattering, the $(1,1)$ and $(2,2)$ elements of $S_{\text {ano }}(\mathbf{q}, E)$ matrix represents the spin up and spin down neutron scattering, respectively. $H$ is the total Hamiltonian for materials but without neutron interaction. For unpolarized neutrons, the neutron spin-averaged anomalous structure factor can be written as
$\frac{1}{2}{\rm{Tr}}{S_{{\rm{ano}}}}({\bf{q}},E) \approx 0$ if we assume $\rho_{\mathbf{q} \uparrow}=\rho_{\mathbf{q} \downarrow}$. Therefore, it indicates that polarized neutron incidence is needed to observe the anomalous scattering, unless the projected magnetism is different for spin up and spin down neutrons. Despite arising as a natural consequence of the interference between two types of interactions, moreover, the Breit-Wigner formula for fast neutrons has been known for decades \cite{yip2014nuclear}, the implication of the anomalous scattering in Eq.\,\eqref{eq:S-ano} for thermal neutron scattering has been elusive.
\begin{figure}
    \centering
 \subfigure[Nuclear-Nuclear scattering]{\includegraphics[width=0.2\textwidth]{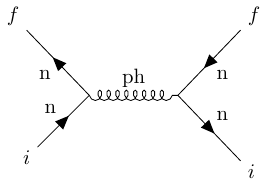}}
 \subfigure[Magnetic-Magnetic scattering]{\includegraphics[width=0.22\textwidth]{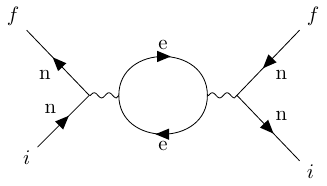}}
 \subfigure[Nuclear-Magnetic scattering]{\includegraphics[width=0.48\textwidth]{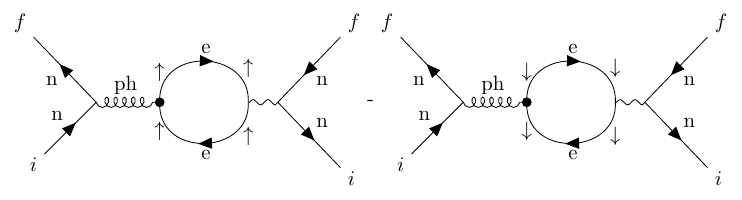}}
    \caption{Figures of Feynman diagram for all Nuclear-Nuclear scattering, Magnetic-Magnetic scattering, and Nuclear-Magnetic scattering.}
    \label{fig:enter-label}
\end{figure}

\textit{Anomalous scattering with electron-phonon interaction} \textemdash
In this section, we show that the presence of the electron-phonon interaction can lead to non-zero anomalous scattering signal. The general electron-phonon interaction Hamiltonian can be written as 
\begin{align}
    H_{\mathrm{e}-\mathrm{ph}}=\frac{N^{1 / 2}}{L^3} \sum_{\mathbf{k} \mathbf{q} \sigma \lambda} g_{\mathbf{q} \lambda} c_{\mathbf{k}+\mathbf{q} \sigma}^{\dagger} c_{\mathbf{k} \sigma}\left(a_{\mathbf{q} \lambda}+a_{-\mathbf{q} \lambda}^{\dagger}\right)
\end{align}
where $\left[a_{\mathbf{q} \lambda}, a_{\mathbf{q}^{\prime} \lambda^{\prime}}^{\dagger}\right]=\delta_{\mathbf{q} \mathbf{q}^{\prime}} \delta_{\lambda \lambda^{\prime}}$ are the phonon operators, $g_{\mathbf{q} \lambda}$ is the electron-phonon coupling constant and is model dependent. For the full materials Hamiltonian $H=H_0+H_{\mathrm{int}}$, defining the anomalous response function in imaginary time as
\begin{align}
& \chi_{\rho M}^z\left(\mathbf{q}, \tau_1-\tau_2\right)=-\mathrm{T}_\tau\left\langle\rho_{\mathbf{q}}\left(\tau_1\right) M_{-\mathbf{q} \perp}^z\left(\tau_2\right)\right\rangle_H 
\nonumber\\
& =-\frac{\mathrm{T}_\tau\left\langle U(\beta) \hat{\rho}_{\mathbf{q}}\left(\tau_1\right) \hat{M}_{-\mathbf{q} \perp}^z\left(\tau_2\right)\right\rangle_{H_0}}{\langle U(\beta)\rangle_{H_0}}
\label{eq:ano-response}
\end{align}
where $U(\beta) \equiv T_{\tau^{\prime}} \exp \left(-\int_0^\beta d \tau^{\prime} \hat{H}_{\text {int }}\left(\tau^{\prime}\right)\right)$ is the interaction picture time evolution operator that evolves with $H_0$, ``$\widehat{}$'' means the operators in interaction picture, i.e. $\hat{A}(\tau) \equiv e^{+\tau H_0} A e^{-\tau H_0}$. The $H_{\mathrm{int}}$ here is a general term that could be electron-phonon interaction but is not necessary to be limited to it. For weak electron-phonon interaction with $H=H_0+H_{\mathrm{e-ph}} = H_{e}+H_{ph}+H_{\mathrm{e-ph}}$, we have $U(\beta) \approx 1-\int_0^\beta d \tau \hat{H}_{\mathrm{e}-\mathrm{ph}}(\tau)$, while for small lattice displacement, we have $\rho_{\mathbf{q}} \approx \bar{b} \sum_l e^{-i \mathbf{q} \cdot \mathbf{R}_l^0}\left(1-i \mathbf{q} \cdot \mathbf{u}\left(\mathbf{R}_l^0\right)\right)$, in which $\bar{b}$ is the averaged nuclear scattering length density, and the lattice displacement for an atom located at $\mathbf{R}_l^0$ can be written as $\mathbf{u}\left(\mathbf{R}_l^0\right)=\sum_{\mathbf{q} \lambda} \sqrt{\frac{\hbar}{2 M N \omega_{\mathbf{q}\lambda} }}\left(a_{\mathbf{q} \lambda}+a_{-\mathbf{q} \lambda}^{\dagger}\right) e^{i \mathbf{q} \cdot \mathbf{R}_l^0} \varepsilon_{\mathbf{q} \lambda}$, then the most general anomalous response function Eq.\,\eqref{eq:ano-response} can be written as (see Supplemental Information C \cite{supp})
\begin{align} \label{eq:suscep-full}
& \chi_{\rho M}^z\left(\mathbf{q}, \tau_1-\tau_2\right)= \nonumber \\
& -\frac{N}{L^3} \bar{b} \int_0^\beta d \tau \sum_{\substack{\mathbf{k}^{\prime} \sigma \nonumber\\
\mathbf{k}^{\prime} \sigma^{\prime} \sigma^{\prime \prime}}} g_{-\mathbf{q}} \sqrt{\frac{\hbar}{2 M \omega_{\mathbf{q}}}}\left(-i \mathbf{q} \cdot \varepsilon_{\mathbf{q}}\right) \mathbf{m}_{\perp \sigma^{\prime} \sigma^{\prime \prime}}^z(-\mathbf{q}) \times \nonumber \\
& \mathrm{T}_\tau\left\langle c_{\mathbf{k}^{\prime}-\mathbf{q} \sigma}^{\dagger}(\tau) c_{\mathbf{k}^{\prime} \sigma}(\tau) c_{\mathbf{k}^{\prime \prime} \sigma^{\prime}}^{\dagger}\left(\tau_2\right) c_{\mathbf{k}^{\prime \prime}-\mathbf{q} \sigma^{\prime \prime}}\left(\tau_2\right)\right\rangle_{H_e} D_{-\mathbf{q}}^{(0)}\left(\tau-\tau_1\right)
\end{align}
in which $D_{-\mathbf{q}}^{(0)}\left(\tau-\tau_1\right)$ is the free-phonon propagator, which is connected to Matsubara-domain phonon propagator as $D_{\mathbf{q}}^{(0)}\left(i \omega_n\right)=\int_0^\beta d \tau e^{+i \omega_n \tau} D_{\mathbf{q}}^{(0)}\left(\tau=\tau_1-\tau_2\right)=-\frac{2 \hbar \omega_{\mathbf{q}}}{\omega_n^2+\left(\hbar \omega_{\mathbf{q}}\right)^2} \quad$. Here we have assumed the single mode phonon and get rid of $\lambda$ without loss of generality. To obtain the anomalous structure factor, we define the Matsubara-frequency domain response function $\chi_{\rho M}^z\left(\mathbf{q}, i \omega_n\right)=\int_0^\beta \chi_{\rho M}^z(\mathbf{q}, \tau) e^{i \omega_n \tau} d \tau$, which is connected to the retarded response function $\chi_{\rho M}^{z, R}(\mathbf{q}, E)$ via analytical continuation $\chi_{\rho M}^{z, R}(\mathbf{q}, E)=\hbar \chi_{\rho M}^z\left(\mathbf{q}, i \omega_n \rightarrow E+i \delta\right)$. Finally, the retarded response function is connected to the anomalous structure factor $S_{\rho M \uparrow /\downarrow}^z(\mathbf{q}, E)$ through the fluctuation-dissipation theorem $S_{\rho M}^z(\mathbf{q}, E)=-2\left[n_B(E)+1\right] \operatorname{Im} \chi_{\rho M}^{z, R}(\mathbf{q}, E)$.

\begin{figure}
    \centering
     \includegraphics[width=0.4\textwidth]{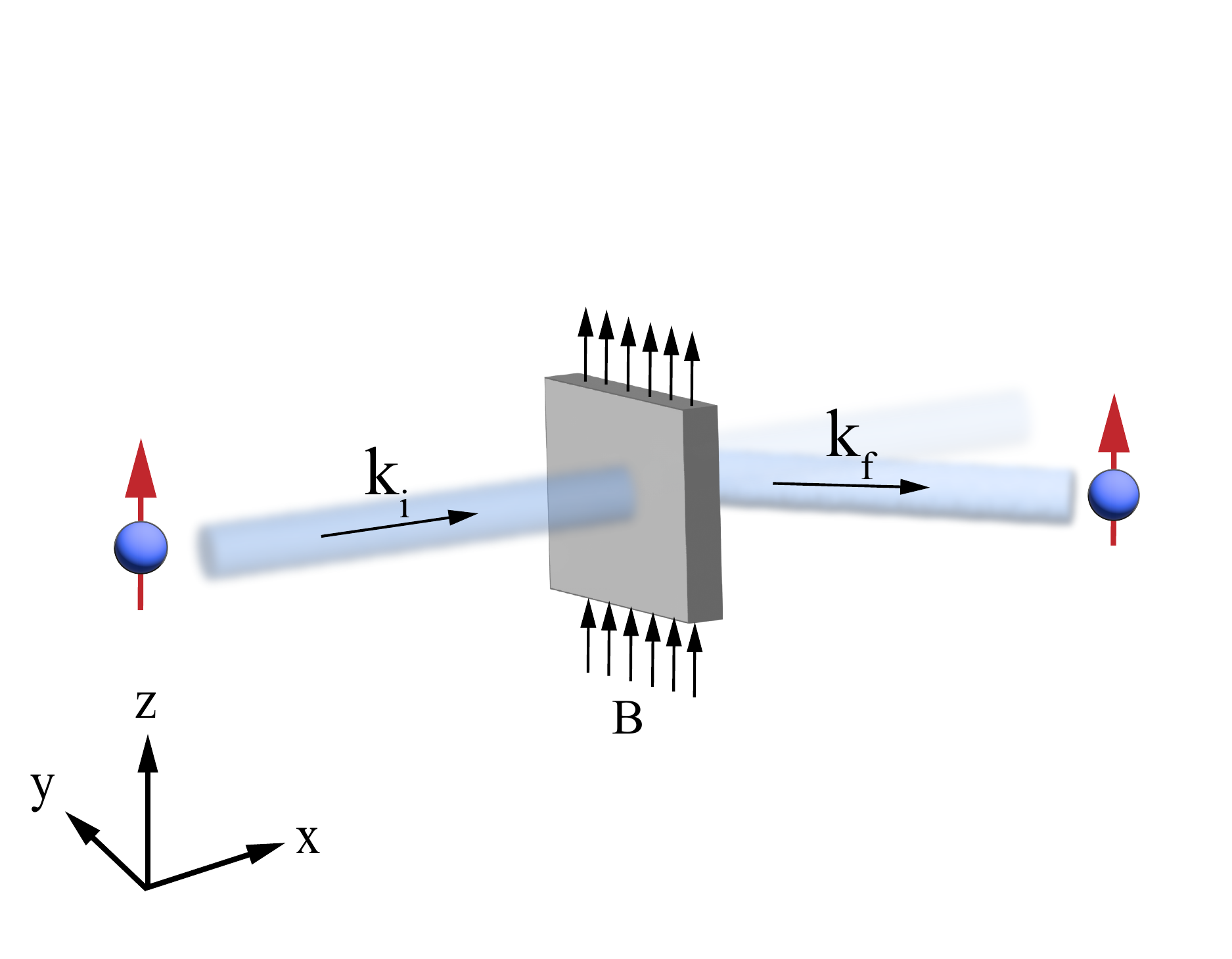} 
    \caption{The illustration diagram of the ANUBIS setup with the polarized neutron scattering technique. This setup is possible to be constructed as it aligns with the commonly used inelastic neutron scattering setup.}
    \label{fig:2}
\end{figure}
To simplify Eq.\,\eqref{eq:suscep-full}, we assume that the spin and momentum degrees of freedom are fully decoupled. We can define spin-dependent electron number density $\rho_{e \mathbf{q} \sigma}=\sum_{\mathbf{k}^{\prime}} c_{\mathbf{k}^{\prime}-\mathbf{q} \sigma}^{\dagger} c_{\mathbf{k}^{\prime} \sigma}$, and the spin-dependent charge density response function $\chi_{\mathbf{q} \sigma}^{e e}\left(i \omega_n\right)=-\int_0^\beta d \tau \mathrm{T}_\tau\left\langle\rho_{e \mathbf{q} \sigma}(\tau) \rho_{e-\mathbf{q} \sigma}\left(\tau_2\right)\right\rangle_{H_e} e^{+i \omega_n \tau}$. This can be written in Matsubara frequency as 
\begin{align}\label{eq:suscep-simple}
\chi_{\rho M}^z\left(\mathbf{q}, i \omega_n\right)= & \frac{N}{L^3} \bar{b} g_{-\mathbf{q}} \sqrt{\frac{\hbar}{2 M \omega_{\mathbf{q}}}}\left(-i \mathbf{q} \cdot \varepsilon_{\mathbf{q}}\right) \nonumber\\
& \times \sum_\sigma \mathbf{m}_{\perp \sigma \sigma}^z(-\mathbf{q}) \chi_{\mathbf{q} \sigma}^{e e}\left(i \omega_n\right) D_{-\mathbf{q}}^{(0)}\left(-i \omega_n\right)
\end{align}
Eq.\,\eqref{eq:suscep-simple} is unusual in the following sense. First, it shows that the anomalous response $\chi_{\rho M}^z\left(\mathbf{q}, i \omega_n\right) \propto \bar{b} g_{-\mathbf{q}}$ is linear in both nuclear scattering length density $\bar{b}$ and electron-phonon interaction $g_{-\mathbf{q}}$, which is due to the interference effect. This starkly contrasts any scattering event; even with lowest order scattering, Fermi's golden rule will give either $|\bar{b}|^2$ or $\left|g_{-q}\right|^2$ dependence. Given the small signal of $g_{-\mathbf{q}}$, the linear dependence could benefit its observation. Second, it is also proportional to the electron charge-density response $\chi_{\mathbf{q} \sigma}^{e e}\left(i \omega_n\right)$ but at phonon wavevector $\mathbf{q}$. Thirdly, the signal is also linearly dependent on the relative angle between the phonon wavevector and phonon polarization, $\left(\mathbf{q} \cdot \varepsilon_{\mathbf{q}}\right)$, indicating that it can either increase (constructive interference) or decrease (destructive interference) the neutron counts. This feature is in contrast to the inelastic phonon scattering signal $\propto\left(\mathbf{q} \cdot \varepsilon_{\mathbf{q}}\right)^2$.

\textit{The ANUBIS signal in homogeneous Fermi gas and semimetals} \textemdash Since neutrons are seldom used in detecting the contribution of the orbital motion of electrons even in the conventional setting, we concentrate on the electron spin degrees of freedom here with the brief discussion of the orbital contribution in the supplementary information C\,\cite{supp}. Notice that ${\bf{m}}_{ \bot ,{\rm{spin}}}^z( - {\bf{q}}) = {\bf{m}}_{ \bot ,{\rm{spin}}}^z( + {\bf{q}}) = \frac{{{\bf{q}} \times {{\bf{s}}_z} \times {\bf{q}}}}{{{q^2}}}$, ${{\bf{s}}_z} = \frac{1}{2}\left( {\begin{array}{*{20}{c}}
1&0\\
0&1
\end{array}} \right)$ is dimensionless in our notation, then we have
\begin{equation}
\begin{split}
    \chi _{\rho M}^z({\bf{q}},i{\omega _n}) = \frac{N}{L^3}\bar b{g_{ - {\bf{q}}}}\sqrt {\frac{\hbar }{{2M{\omega _{\bf{q}}}}}} \left( { - i{\bf{q}} \cdot {\varepsilon _{\bf{q}}}} \right)\\
 \times {\bf{m}}_{ \bot ,{\rm{spin}} \uparrow  \uparrow }^z({\bf{q}})\left( {\chi _{{\bf{q}} \uparrow }^{ee}(i{\omega _n}) - \chi _{{\bf{q}} \downarrow }^{ee}(i{\omega _n})} \right)D_{ - {\bf{q}}}^{(0)}( - i{\omega _n})
 \label{eq:ano-MResponse}
\end{split}
\end{equation}
The polarization part is further simplified with the help of weak $B$ field approximation and the Lindhard formula under the static approximation\,\cite{coleman2015introduction}. This approximation is valid since the phonon energy measured by neutron scattering is small relative to the Fermi energy, while the $q$ can be arbitrary in the Brillouin zone. We have $\chi _{{\bf{q}} \uparrow }^{ee}(i{\omega _n}) - \chi _{{\bf{q}} \downarrow }^{ee}(i{\omega _n}) = 2{\mu _B}{B_z}\frac{{m_e^2L^3}}{{q{\pi ^2}{\hbar ^4}}}\ln \left| {\frac{{q + 2{k_F}}}{{q - 2{k_F}}}} \right|$ for the homogeneous electron gas. Assuming the $\bf{q}$ is within the $xy$ plane (Fig. \ref{fig:2}), we have ${\bf{m}}_{ \bot ,{\rm{spin}} \uparrow  \uparrow }^z({\bf{q}}) = \frac{1}{2}$, the final result will be
\begin{equation}
\begin{split}
\chi _{\rho M}^z({\bf{q}},i{\omega _n}) =N\bar b{g_{ - {\bf{q}}}}\sqrt {\frac{\hbar }{{2M{\omega _{\bf{q}}}}}} \left( { - i{\bf{q}} \cdot {\varepsilon _{\bf{q}}}} \right)D_{ - {\bf{q}}}^{(0)}( - i{\omega _n})\\
 \times \frac{{m_e^2}}{{q{\pi ^2}{\hbar ^4}}}\ln \left| {\frac{{q + 2{k_F}}}{{q - 2{k_F}}}} \right|{\mu _B}{B_z}
\end{split}
\end{equation}

The complete formulas with the full momentum and energy dependence for normal metal and topological semimetal are also derived and can be consulted within Supplemental Information D \cite{supp}.


\begin{figure}
    \centering
     \includegraphics[width=0.5\textwidth]{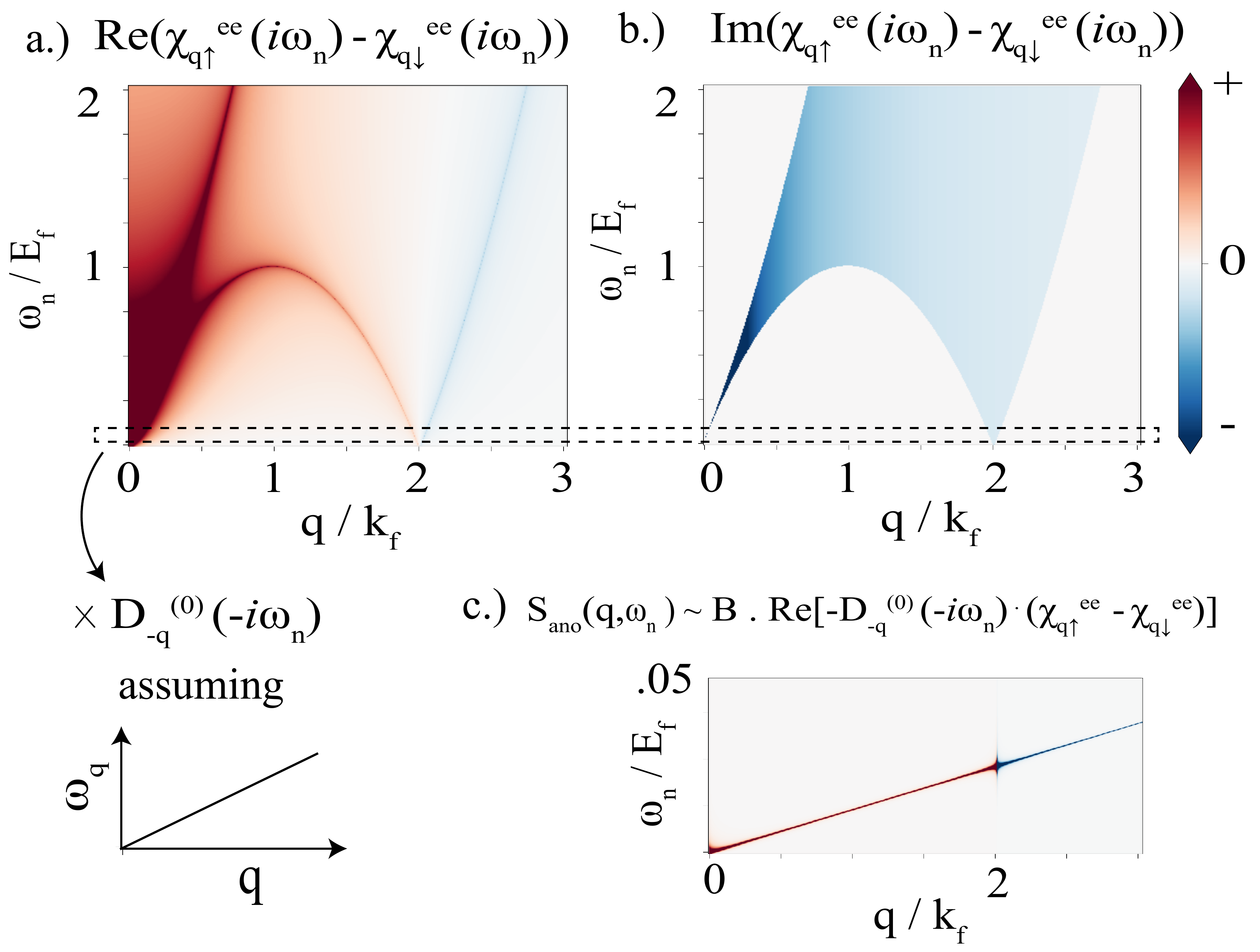} 
    \caption{The numerical results of the anomalous response for homogeneous electron gas. Figure (a) and Figure (b) are the real part and the imaginary part of the difference between the spin-dependent charge polarizations. Figure (c) sketches the anomalous structure factor under the finite magnetic field associated with the assumed linear-dispersive acoustic phonons reflected in the free phonon propagator. The x-axis and y-axis are the unitless momentum and energy, and all the linear parameters' contributions including the magnetic field are all normalized to 1.}
    \label{fig:nr1}
\end{figure}

\textit{The connection to dielectric function} \textemdash To see the link of Eq.\,\eqref{eq:suscep-simple} and dielectric function, we focus on deformation potential scattering, where $g_{\mathbf{q}}=i e \sqrt{\frac{\hbar}{2 M \omega_{\mathbf{q}}}}\left(\mathbf{q} \cdot \varepsilon_{\mathbf{q}}\right) Z V_{\mathbf{q}}$, $V_{\mathbf{q}}=\frac{4 \pi e}{q^2+k_s^2}=\frac{4 \pi e}{q^2}\left(\mathrm{k}_s=0\right)$. This is valid for the acoustic phonons. In this situation, defining the spin-dependent dielectric function as $\varepsilon_\sigma^{-1}\left(\mathbf{q}, i \omega_n\right)=1+\frac{1}{L^3} e V_{\mathbf{q}} \chi_{\mathbf{q} \sigma}^{e e}\left(i \omega_n\right)$, Eq.\,\eqref{eq:suscep-simple} can be further simplified as
\begin{equation}
\begin{split}
\chi_{\rho M}^z\left(\mathbf{q}, i \omega_n\right)= & C_{\mathbf{q}} D_{-\mathbf{q}}^{(0)}\left(-i \omega_n\right) \sum_\sigma \mathbf{m}_{\perp \sigma \sigma}^z(-\mathbf{q}) \\
& \times \left(\frac{1}{\varepsilon_\sigma\left(\mathbf{q}, i \omega_n\right)}-1\right) 
\label{eq:Dielec}
\end{split}
\end{equation}
where $\quad C_{\mathbf{q}} \equiv \frac{\hbar N Z}{2 M \omega_{\mathbf{q}}} \bar{b}\left(\mathbf{q} \cdot \varepsilon_{\mathbf{q}}\right)^2$ is a constant that is determined by the type of atomic nucleus and phonon properties. Eqs.\,\eqref{eq:suscep-simple} and \eqref{eq:Dielec} are the main results of this work.

As a prominent example, we calculate the anomalous scattering from a homogeneous electron gas with Fermi wave vector $k_F$ and Fermi energy $\varepsilon_F$. We consider a prototype case with only electronic spin degrees of freedom $\mathbf{s}=\mathbf{s}_z$ and $\mathbf{q}$ is within the scattering $x y$ plane, then Eq.\,\eqref{eq:Dielec} can further be simplified as 
\begin{align}
    \chi_{\rho M}^z\left(\mathbf{q}, i \omega_n\right)=\frac{1}{2} C_{\mathbf{q}} D_{-\mathbf{q}}^{(0)}\left(-i \omega_n\right) \sum_{\sigma= \pm} \sigma \frac{1}{\varepsilon_\sigma\left(\mathbf{q}, i \omega_n\right)}
\end{align}
which gives zero without an external magnetic field. If external magnetic is applied, a nontrivial result would emerge. Since neutron scattering measures phonon $\sim \mathrm{meV}$ energy range but at arbitrary wavevector of $\mathbf{q}$, we assume a static dielectric function from Thomas-Fermi dielectric
screening, $\quad \varepsilon_\sigma\left(\mathbf{q}, i \omega_n\right) \approx \varepsilon_\sigma(\mathbf{q}, 0)=1+\frac{q_{T F \sigma}^2}{q^2} \quad, \quad$ where $q_{T F \sigma}=\left(\frac{4 k_{F \sigma}}{\pi a_0}\right)^{\frac{1}{2}}$ is the electron-spin-dependent screening wavevector, and $a_0$ is the Bohr radius. Under external Zeeman field, we have $\varepsilon_{F \sigma}=\frac{\hbar^2 k_{F \sigma}^2}{2 m_e}=\varepsilon_F+\sigma \mu_B B_z$. For weak magnetic field, we have the following expression $\sum_{\sigma= \pm} \sigma \frac{1}{\varepsilon_\sigma\left(\mathbf{q}, i \omega_n\right)}=\frac{{4{k_F}\pi {a_0}{q^2}{\mu _B}{B_z}}}{{{{\left( {\pi {a_0}{q^2} + 4{k_F}} \right)}^2}{\varepsilon _F}}}$,
which indicates that the anomalous scattering signal is proportional to the magnetic field $B$. The schematic configuration is shown in Fig.\,\ref{fig:2}. Through the fluctuation-dissipation theorem, finally, we should have:
\begin{equation}
\begin{split}
S_{\rho M}^z({\bf{q}},E) 
&=
 {C_{\bf{q}}}\left[ {{n_B}(E) + 1} \right]\frac{{4{\pi ^2}{a_0}{q^2}{k_F}\hbar }}{{{{\left( {\pi {a_0}{q^2} + 4{k_F}} \right)}^2}{\varepsilon _F}}}\\&\times \left[ {\delta \left( {E - \hbar {\omega _{ - {\bf{q}}}}} \right) + \delta \left( {E + \hbar {\omega _{ - {\bf{q}}}}} \right)} \right]{\mu _B}{B_z}
 \label{eq:dielecS}
\end{split}
\end{equation}
Eq.\,\eqref{eq:dielecS} characterize the contributions of both electron and phonon. 

\textit{Signal strength w.r.t. phonon scattering} \textemdash To compare the signal strength of ANUBIS, the structure factor of the phonon scattering is written as (see Supplemental Information B \cite{supp}) $S_{\rho\rho}({\bf{q}},E) =  - 2\left[ {{n_B}(E) +1}\right]\Im\chi^{R}_{{\rho\rho}} ({\bf{q}},E)$ and $\chi _{\rho \rho }^R({\bf{q}},E) = \hbar {\chi _{\rho \rho }}({\bf{q}},i{\omega _n} \to E + i\delta )$, where:
\begin{equation}
\begin{split}
{\chi _{\rho \rho }}({\bf{q}},i{\omega _n}) &=  \frac{{\hbar N{{\bar b}^2}}}{{2M\sqrt {{\omega _{\bf{q}}}{\omega _{ - {\bf{q}}}}} }}\left( {{\bf{q}} \cdot {\varepsilon _{\bf{q}}}} \right)\left( {{\bf{q}} \cdot {\varepsilon _{ - {\bf{q}}}}} \right)D_{\bf{q}}^{(0)}(i{\omega _n})
\end{split}
\end{equation}
Here we have taken the low temperature limit to ignore the Debye-Waller factor. As a result, we get the quotient to compare the signal strength for electron gas with Eq.\,\eqref{eq:dielecS}:
\begin{equation}
\begin{split}
\frac{I_{ANUBIS}}{I_{phonon}}=\frac{2\gamma r_e S_{\rho M }^z}{S_{\rho\rho}} = 
\gamma {r_e}\frac{Z}{{\bar b}}{\frac{{4{k_F}\pi {a_0}{q^2}{\mu _B}{B_z}}}{{{{\left( {\pi {a_0}{q^2} + 4{k_F}} \right)}^2}{\varepsilon _F}}}} 
\end{split}
\end{equation}
Here we have already considered the symmetry of the dynamic matrix to simplify the expression. To estimate the signal strength for the real system, we take the estimation for these parameters: $\gamma = 1.913$, $r_e = 2.8 \times 10fm$, $q = 10^{10} m^{-1}$, $B_z = 10 T$, Bohr radius $a_0=5.29 \times 10^{-11}m$  Bohr magneton $\sim 5 \times 10^{-5} eV/T$, and make use of the detailed materials parameters, we conclude the signal strength of ANUBIS is around $10^{-4}$ to $10^{-3}$ based on the estimation of these arbitrary selected metal system in Table \ref{table:estimate}.

\todo{Despite the small ANUBIS signal, it is worthwhile mentioning that the the phonon selection rule may cancel out the phonon intensity, while presence  of electron-phonon interaction leads to less-restrictive selection rule and supports the experimental observation.}

\begin{table}[h!]
\centering
\begin{tabular}{||c||c||c||} 
\hline
Materials& Parameters  & Signal strength\\ 
\hline\hline
Cu(fcc) &$Z= 29$, $\varepsilon _F =7eV$, & $2.585 \times 10^{-4}$ \\&$\bar b \approx 7.718fm$ &	\\
\hline
Ag(fcc) &$Z= 47$, $\varepsilon _F =5.49eV$, & $7.405 \times 10^{-4}$ \\&$\bar b \approx 5.922fm$ &	\\
\hline
Au(bcc) & $Z= 79$, $\varepsilon _F =5.53eV$, & $9.574 \times 10^{-4}$ \\&$\bar b \approx 7.63fm$ &	\\
\hline
Nb(bcc) &$Z= 41$, $\varepsilon _F =5.32eV$,& $5.639 \times 10^{-4}$ \\&$\bar b \approx 7.054fm$ &	\\
\hline
Mg(hcp) &$Z= 12$, $\varepsilon _F =7.08eV$, & $1.514 \times 10^{-4}$ \\&$\bar b \approx 5.375fm$ &	\\
\hline
Tl(hcp) &$Z= 81$, $\varepsilon _F =8.15eV$,& $5.233 \times 10^{-4}$ \\&$\bar b \approx 8.776fm$ &	\\
\hline
Sn(bct) &$Z= 50$, $\varepsilon _F =10.2eV$,& $3.411 \times 10^{-4}$ \\&$\bar b \approx 6.225fm$ &	\\
\hline
Ga(orthorhombic) &$Z= 31$, $\varepsilon _F =10.4eV$,& $1.762 \times 10^{-4}$ \\&$\bar b \approx 7.288fm$ &	\\
\hline
Sb(rhombohedral) &$Z= 51$, $\varepsilon _F =10.9eV$,& $3.567 \times 10^{-4}$ \\&$\bar b \approx 5.57fm$ &	\\
\hline
\hline
\end{tabular}
\caption{The table of order-of-magnitude w.r.t. phonon scattering for several common metal systems.}
\label{table:estimate}
\end{table}

\textit{Numerical Results} \textemdash The response function Eq.\,\eqref{eq:ano-MResponse} has also been numerically calculated for the homogeneous electron gas with momentum and energy dependence in Fig.\,\ref{fig:nr1}. \todo{For linear-dispersive acoustic phonons, the signal keeps positive in the range of $q<2k_f$ due to constructive interference, and flips to negative when $q>2k_f$. The $q=2k_f$ coincides with the condition of Kohn anomaly, indicating that ANUBIS signal can be amplified at Kohn anomaly point by varying magnetic field.} The detailed and complete derivation should refer to Supplemental Information D \cite{supp}. As we noticed the numerical calculations of the electron gas and semimetal behave very differently, which indicates the ANUBIS signal largely reflects the electronic energy dispersion relation.

To summarize, we propose an experimental scheme with Nuclear-Magnetic Interference Spectroscopy to acquire information about the electron properties through polarized neutron scattering. The surprising existence of the linear dependence of the electron-phonon interaction term within the structure factor reveals a new routine to measure the electronic properties with neutron scattering. Moreover, the connection with the momentum and energy-resolved dielectric function is discussed. The order-of-magnitude estimation indicates the anomalous scattering signal is around $10^{-4}$ to $10^{-3}$ with respect to phonon scattering, which exactly satisfies the least required signal strength of the capability for the next-generation neutron facility such as STS and ESS. The whole results obtained have indicated the proposed ANUBIS has strong potential to be a powerful technique for measuring crucial electronic properties. Future experiments configuration based on this proposal will benefit various studies of the phenomena associated with significant participated electron-phonon coupling, such as unconventional superconductivity \cite{lanzara2001evidence}, flat-band system \cite{feng2020interplay,drucker2024incipient}, transport \cite{tong2019comprehensive}, laser-material processing \cite{PhysRevB.77.075133}, charge density waves \cite{lian2020ultrafast,sayers2020coherent} like AV3Sb5 \cite{luo2022electronic}, localized state \cite{atta2004electron}, thermoelectric performance \cite{Heng2012,Lizhong2022}, etc. With improved time-resolved power, it might even be possible to combine the ultrafast pump-probe spectroscopy design to establish the time-resolved setup to reflect materials' electron-phonon interaction\cite{sentef2017light,hu2022tracking} or during the phase transition \cite{gidopoulos2022enhanced}. We expect ANUBIS will also open up the opportunity and broaden the application to study the electron-phonon interaction under extreme conditions such as irradiation \cite{zarkadoula2017effects} and high-pressure conditions \cite{john2004electron,PhysRevB.96.100502} with the advanced \textit{in-situ} techniques. \todo{Finally, we point out that ANUBIS effect can probe the spin-phonon interaction \cite{hong2024phonon} and chiral phonon \cite{zhu2018} beyond the electron-phonon interaction.}

The authors thank Jeffrey W. Lynn, Rebecca Dally and Yao Wang for the insightful discussions. CF acknowledges support from the US Department of Energy (DOE), Office of Science (SC), Basic Energy Sciences (BES), Award No. DE-SC0020148. PS acknowledges support from DOE Award No. DE-SC0021940. CL acknowledges the support from DOE DE-SC0023874, WX thanks support from DOE DE-SC0023648. AB acknowledges the support from the National Science Foundation (NSF) Convergence Accelerator Award No. 2235945. ML is partially supported by the Class of 1947 Career Development Chair and support from R Wachnik.


\nocite{*}

\bibliography{references}

\pagebreak
\widetext
\center

\setcounter{page}{1}

\large
Supplemental Information

\vspace{0.5in}

\textbf{Anomalous Neutron Nuclear-Magnetic Interference Spectroscopy}

\vspace{0.5in}


\normalsize
\raggedright
This supplementary file contains the following elements:\\
\textbf{Supplementary Text with Sections A to D}\\

\makeatletter
\renewcommand \thesection{S\@arabic\c@section}
\renewcommand\thetable{S\@arabic\c@table}
\renewcommand \thefigure{S\@arabic\c@figure}
\renewcommand \theequation{S\@arabic\c@equation}
\setcounter{figure}{0}
\setcounter{equation}{0}
\setcounter{figure}{0}
\setcounter{table}{0}
\makeatother

\section{A. Anomalous dynamical structure factor with quantum many-body theory}
When both neutron nuclear interaction ${H_{I,{\rm{nuc}}}}$ and neutron magnetic interaction ${H_{I,{\rm{mag}}}}$ co-exist, the total neutron-matter interaction Hamiltonian can be written as 
\begin{equation}
\begin{split}
    {H_I} &= \frac{g}{{{L^3}}}\sum\limits_{{\bf{kq}}} {\Psi _{N{\bf{k}} + {\bf{q}}}^ + \left( {{\rho _{\bf{q}}} - \gamma {r_e}{{\bf{\sigma }}_N} \cdot {{\bf{M}}_{ \bot {\bf{q}}}}} \right){\Psi _{N{\bf{k}}}}} \\
 &= \frac{g}{{{L^3}}}\sum\limits_{{\bf{kq}}} {\left( {\begin{array}{*{20}{c}}
{\Psi _{N{\bf{k}} + {\bf{q}} \uparrow }^ + }&{\Psi _{N{\bf{k}} + {\bf{q}} \downarrow }^ + }
\end{array}} \right)\left( {\begin{array}{*{20}{c}}
{{\rho _ \uparrow }_{\bf{q}} - \gamma {r_e}M_{ \bot {\bf{q}}}^{z}}&{ - \gamma {r_e}M_{ \bot {\bf{q}}}^{x} + i\gamma {r_e}M_{ \bot {\bf{q}}}^{y}}\\
{ - \gamma {r_e}M_{ \bot {\bf{q}}}^{x} - i\gamma {r_e}M_{ \bot {\bf{q}}}^{y}}&{{\rho _{ \downarrow {\bf{q}}}} + \gamma {r_e}M_{ \bot {\bf{q}}}^{z}}
\end{array}} \right)\left( {\begin{array}{*{20}{c}}
{{\Psi _{N{\bf{k}} \uparrow }}}\\
{{\Psi _{N{\bf{k}} \downarrow }}}
\end{array}} \right)} 
\end{split}
\end{equation}

To calculate the neutron scattering cross-sections, we can write down the neutron interacting self-energy matrix in Matsubara domain: 
\begin{equation}
\Sigma _N^{y}({\bf{k}},i{p_m}) =  - {\left( {\frac{g}{{{L^3}}}} \right)^2}\frac{1}{\beta }\sum\limits_{{\bf{q}}{\omega _n}} {G_N^{(0)}({\bf{k}} - {\bf{q}},i{p_m} - i{\omega _n})X({\bf{q}},i{\omega _n})} 
\end{equation}

where the non-interacting neutron propagator matrix for the two neutron spin component 
\begin{equation}
G_N^{(0)}({\bf{k}} - {\bf{q}},i{p_m} - i{\omega _n}) = \left( {\begin{array}{*{20}{c}}
{\frac{1}{{i{p_m} - i{\omega _n} - {E_{N{\bf{k}} - {\bf{q}}}}}}}&0\\
0&{\frac{1}{{i{p_m} - i{\omega _n} - {E_{N{\bf{k}} - {\bf{q}}}}}}}
\end{array}} \right)
\end{equation}

and the total Matsubara response function $X^{\sigma_f\sigma_i}({\bf{q}},i{\omega _n})$ is defined as 
\begin{equation}
    \begin{split}
X^{\sigma_f \sigma_i}({\bf{q}},i{\omega _n}) &= \int\limits_0^\beta  {X({\bf{q}},\tau ){e^{i{\omega _n}\tau }}d\tau } \\
X^{\sigma_f \sigma_i}({\bf{q}},{\tau _1} - {\tau _2}) &=  
- {{\rm{T}}_\tau }{\left\langle {\left( {\rho_{\bf{q}} ({\tau _1})\bra{\sigma_i}\ket{\sigma_f} - \gamma {r_e}\bra{\sigma_i}\bm{\sigma_N}\ket{\sigma_f} \cdot {{\bf{M}}_{ \bot {\bf{q}}}}({\tau _1})} \right)\left( {{\rho_{ - {\bf{q}}}}({\tau _2})\bra{\sigma_f}\ket{\sigma_i} - \gamma {r_e}\bra{\sigma_f}\bm{\sigma_N}\ket{\sigma_i} \cdot {{\bf{M}}_{ \bot  - {\bf{q}}}}({\tau _2})} \right)} \right\rangle _H}\\
 &= \frac{1}{\beta }\sum\limits_n {X^{\sigma_f \sigma_i}({\bf{q}},i{\omega _n}){e^{ - i{\omega _n}({\tau _1} - {\tau _2})}}} 
    \end{split}
\end{equation}

where $\sigma_i, \sigma_f$ is the initial and finial spin polarization of the neutron. We can then define magnetic structure factor in a matrix form $S(\bm{q}, E) = \begin{pmatrix}
    S^{\uparrow\uparrow}(\bm{q}, E) & S^{\uparrow\downarrow} 
    (\bm{q}, E) \\
     S^{\downarrow\uparrow}(\bm{q}, E) & S^{\downarrow\downarrow} 
    (\bm{q}, E)
\end{pmatrix}$ 
where the diagonal part and the off-diagonal part of the matrix reflect the spin-flip part and the non-spin-flip part of the structure factor, respectively. For the nuclear and magnetic potential, the magnetic structure factor can be written as 
\begin{equation}
    \begin{split}
{S_{{\rm{tot}}}}({\bf{q}},E) &= {S_{\rho \rho }}({\bf{q}},E) + {(\gamma {r_e})^2}{S_{MM}}({\bf{q}},E) - 2\gamma {r_e}{S_{{\rm{ano}}}}({\bf{q}},E)\\
 &= \underbrace {\left( {\begin{array}{*{20}{c}}
{{S_{\rho \rho , \uparrow  \uparrow }}({\bf{q}},E)}&0\\
0&{{S_{\rho \rho , \downarrow  \downarrow }}({\bf{q}},E)}
\end{array}} \right)}_{{\rm{nuclear\, scattering}}} + \underbrace {{{(\gamma {r_e})}^2}  { \begin{pmatrix}
    S_{M^z M^z}(\bm{q}, E) & S_{M^- M^+}(\bm{q}, E)\\
    S_{M^+ M^-}(\bm{q}, E) & S_{M^z M^z}(\bm{q}, E)
    \end{pmatrix}} }_{{\rm{magnetic\, scattering}}}\\
 &- \underbrace {2\gamma {r_e} \left(\begin{array}{cc}
S_{\rho M \uparrow}^z(\mathbf{q}, E) & 0 \\
0 & -S_{\rho M \downarrow}^z(\mathbf{q}, E)
\end{array}\right)}_{{\rm{anomalous\,scattering}}}
    \end{split}
    \label{eq:S-tot}
\end{equation}

in which the nuclear and magnetic dynamical structure factors are well known and defined as Eq.\,\eqref{eq:S-rho} and Eq.\,\eqref{eq:S-mag}. The anomalous dynamical structure factor ${S_{{\rm{ano}}}}({\bf{q}},E)$ arises from the crossing term between nuclear and magnetic scattering as 
\begin{equation}
S_{\rho M \uparrow / \downarrow }^j({\bf{q}},E) = \int\limits_{ - \infty }^{ + \infty } {{{\left\langle {{\rho _{{\bf{q}} \uparrow / \downarrow }}(t)M_{ - {\bf{q}} \bot }^j(0)} \right\rangle }_H}} {e^{iEt/\hbar }}dt
\end{equation}
which is beyond the decades-long bi-partite of the nuclear dynamical structure factor and the magnetic dynamical structure factor as normally done in Eq.\,\eqref{eq:S-rho} and Eq.\,\eqref{eq:S-mag}. In Eq.\,\eqref{eq:S-tot}, each matrix element indicates the corresponding structure factor event for neutron spin non-flip (diagonal) and spin flip (off-diagonal) scatterings for up and down spins, respectively.

\section{B. Nuclear and Magnetic dynamic structure factor with quantum many-body theory}
Under the Fermi pseudopotential approximation, the short-range neutron-nuclear interaction Hamiltonian can be written as: 
\begin{equation}
{H_{I,{\rm{nuc}}}} = g\int {{d^3}{\bf{r}}\Psi _N^ + ({\bf{r}}){\Psi _N}({\bf{r}})\rho ({\bf{r}})} 
\label{intHrho}
\end{equation}
where a local interaction is assumed with interaction strength $g = \frac{{2\pi {\hbar ^2}}}{{{m_N}}}$, neutron field operator ${\Psi _N}({\bf{r}})$. $\rho ({\bf{r}}) = \sum\limits_l {{b_l}\delta ({\bf{r}} - {{\bf{R}}_l})} $ is the nuclear density operator by with neutron nuclear scattering length $b_l$, the summation is over all nuclei located at $\bf{R_l}$, whose Fourier transform is given by $\rho ({\bf{q}}) = \sum\limits_l {{b_l}{e^{ - i{\bf{q}} \cdot {{\bf{R}}_l}}}}$.Without loss of generality, the definition here of $\rho(\bf{q})$ is different from the main text where neutron spins are considered and
$\rho_{\bf{q}}$ is a 2x2 matrix, but we don’t need to deal with neutron spins for simplicity. Using the Fourier-transformed neutron field operator ${\Psi _{N{\bf{k}}}} = {L^{ - 3/2}}\int {{d^3}{\bf{r}}{e^{ - i{\bf{k}} \cdot {\bf{r}}}}{\Psi _N}({\bf{r}})} $, the neutron–matter interaction Hamiltonian Eq.\,\eqref{intHrho} is rewritten as
\begin{equation}
{H_{I,{\rm{nuc}}}} = \frac{g}{{{L^3}}}\sum\limits_{{\bf{kq}}} {\Psi _{N{\bf{k}} + {\bf{q}}}^ + {\Psi _{N{\bf{k}}}}\rho ({\bf{q}})}
\end{equation}
Also (spinless) non-interacting neutron Hamiltonian can be written as 
\begin{equation}
{H_N} = \sum\limits_{\bf{k}} {{E_{N{\bf{k}}}}\Psi _{N{\bf{k}}}^ + {\Psi _{N{\bf{k}}}}}  = \sum\limits_{\bf{k}} {\frac{{{\hbar ^2}{k^2}}}{{2{m_N}}}\Psi _{N{\bf{k}}}^ + {\Psi _{N{\bf{k}}}}} 
\end{equation}
To proceed, we calculate neutron self-energy. The one-loop self-energy $\Sigma _N^{y}({\bf{k}},i{p_n})$ can be written as
\begin{equation}
\Sigma _N^{y}({\bf{k}},i{p_m}) =  - \frac{{{g^2}}}{{{\beta ^1}{L^6}}}\sum\limits_{n{\bf{q}}} {G_N^{(0)}({\bf{k}} - {\bf{q}},i{p_m} - i{\omega _n})\chi ({\bf{q}},i{\omega _n})} 
\label{selfenergyrho}
\end{equation}
where ${\omega _n} = \frac{{2\pi n}}{\beta },n = 0, \pm 1, \pm 2,...$, ${p_m} = \frac{{(2\pi  + 1)m}}{\beta },m = 0, \pm 1, \pm 2,...$ are Bosonic and fermionic Matsubara frequencies. The free-neutron Matsubara Green’s function can be written as 
\begin{equation}
G_N^{(0)}({\bf{k}} - {\bf{q}},i{p_m} - i{\omega _n}) = \frac{1}{{i{p_m} - i{\omega _n} - {E_{N{\bf{k}} - {\bf{q}}}}}}
\end{equation}
The nuclear density-density response function ${\chi _{\rho \rho }}({\bf{q}},i{\omega _n})$, 
\begin{equation}
\begin{split}
{\chi _{\rho \rho }}({\bf{q}},i{\omega _n})& = \int\limits_0^\beta  {\chi ({\bf{q}},\tau  \equiv {\tau _1} - {\tau _2}){e^{i{\omega _n}\tau }}d\tau } \\
{\chi _{\rho \rho }}({\bf{q}},{\tau _1} - {\tau _2}) &=  - {{\rm{T}}_\tau }{\left\langle {\left( {\rho ({\bf{q}}{\tau _1}) - \left\langle {\rho ({\bf{q}}{\tau _1})} \right\rangle } \right)\left( {\rho ( - {\bf{q}},{\tau _2}) - \left\langle {\rho ( - {\bf{q}},{\tau _2})} \right\rangle } \right)} \right\rangle _H}\\
 &= \frac{1}{\beta }\sum\limits_n {\chi ({\bf{q}},i{\omega _n}){e^{ - i{\omega _n}({\tau _1} - {\tau _2})}}} 
\end{split}
\end{equation}
in which ${\chi _{\rho \rho }}({\bf{q}},{\tau _1} - {\tau _2})$ is the density-density response function in “interaction picture” in imaginary time, where the interaction picture means the average is ${\left\langle {} \right\rangle _H}$, referring to the full materials Hamiltonian H but not the term with ${H_{I,{\rm{nuc}}}}$. 
To compare the signal strength of ANUBIS, we derive the structure factor of nuclear scattering. Assume $\left\langle {\rho ({\bf{q}}{\tau _1})} \right\rangle = \left\langle {\rho ( - {\bf{q}},{\tau _2})} \right\rangle = 0$, we have
\begin{equation}
\begin{split}
    {\chi _{\rho \rho }}({\bf{q}},{\tau _1} - {\tau _2}) &=  - {{\rm{T}}_\tau }{\left\langle {\left( {\rho ({\bf{q}}{\tau _1}) - \left\langle {\rho ({\bf{q}}{\tau _1})} \right\rangle } \right)\left( {\rho ( - {\bf{q}},{\tau _2}) - \left\langle {\rho ( - {\bf{q}},{\tau _2})} \right\rangle } \right)} \right\rangle _H} \\
    &\approx \sum\nolimits_{l,l' = 1}^N {{b_l}{e^{ - i{\bf{q}} \cdot {\bf{R}}_l^0}}{b_{l'}}{e^{ + i{\bf{q}} \cdot {\bf{R}}_{l'}^0}}{{\rm{T}}_\tau }{{\left\langle {( - i){\bf{q}} \cdot {{{\bf{\hat u}}}_l}({\tau _1})( - i){\bf{q}} \cdot {{{\bf{\hat u}}}_{l'}}({\tau _2})} \right\rangle }_H}} 
\end{split}
\end{equation}
To simplify the discussion, we take the low-temperature limit to ignore the Debye-Waller factor. Take use of that $\sum\limits_{{\bf{q}}'} {\sum\nolimits_{l,l' = 1}^N {{b_l}{e^{ - i{\bf{q}} \cdot {\bf{R}}_l^0}}{e^{i{\bf{q'}} \cdot {\bf{R}}_l^0}}} }  = N\bar b\sum\limits_{{\bf{q'}}} {{\delta _{{\bf{qq}}'}}}$ and $\sum\limits_{{\bf{q''}}} {\sum\nolimits_{l,l' = 1}^N {{b_l}{e^{ + i{\bf{q}} \cdot {\bf{R}}_l^0}}{e^{i{\bf{q''}} \cdot {\bf{R}}_l^0}}} }  = N\bar b\sum\limits_{{\bf{q''}}} {{\delta _{{\bf{q}}\left( { - {\bf{q''}}} \right)}}}$, here $\bar b$ is the average nuclear scattering length density. Besides, we contract the phonon Green’s function, finally we have
\begin{equation}
    \begin{split}
\chi _{\rho \rho }^{}({\bf{q}},{\tau _1} - {\tau _2}) &= \frac{{\hbar N{{\bar b}^2}}}{{2M\sqrt {{\omega _{\bf{q}}}{\omega _{ - {\bf{q}}}}} }}\left( {{\bf{q}} \cdot {\varepsilon _{\bf{q}}}} \right)\left( {{\bf{q}} \cdot {\varepsilon _{ - {\bf{q}}}}} \right)D_{\bf{q}}^{(0)}({\tau _1} - {\tau _2})\\
{\chi _{\rho \rho }}({\bf{q}},i{\omega _n}) &= \int\limits_0^\beta  {\chi ({\bf{q}},\tau  \equiv {\tau _1} - {\tau _2}){e^{i{\omega _n}\tau }}d\tau }  = \frac{{\hbar N{{\bar b}^2}}}{{2M\sqrt {{\omega _{\bf{q}}}{\omega _{ - {\bf{q}}}}} }}\left( {{\bf{q}} \cdot {\varepsilon _{\bf{q}}}} \right)\left( {{\bf{q}} \cdot {\varepsilon _{ - {\bf{q}}}}} \right)D_{\bf{q}}^{(0)}(i{\omega _n})
    \end{split}
\end{equation}
Now if we define the retarded response function w.r.t. full materials Hamiltonian H 
\begin{equation}
\chi _{\rho \rho }^R({\bf{q}},{t_1} - {t_2}) =  - i\theta ({t_1} - {t_2}){\left\langle {\left[ {\rho ({\bf{q}},{t_1}),\rho ( - {\bf{q}},{t_2})} \right]} \right\rangle _H}
\end{equation}
Then, we can show that we have the Fourier transform
\begin{equation}
\chi _{\rho \rho }^R({\bf{q}},E) = \int\limits_{ - \infty }^{ + \infty } {\chi _{\rho \rho }^R({\bf{q}},{t_1} - {t_2}){e^{iE({t_1} - {t_2})/\hbar }}d{t_1}} 
\end{equation}
Then, we have the central relation linking Matsubara response function to retarded response function as: 
\begin{equation}
\chi _{\rho \rho }^R({\bf{q}},E) = \hbar {\chi _{\rho \rho }}({\bf{q}},i{\omega _n} \to E + i\delta )
\end{equation}
The retarded response function can be further written in terms of the spectral function 
\begin{equation}
 \chi _{\rho \rho }^R({\bf{q}},i{\omega _n}) = \int\limits_{ - \infty }^{ + \infty } {\frac{{{A_{\rho \rho }}({\bf{q}},E)}}{{i{\omega _n} - E + i{0^ + }}}\frac{{dE}}{{2\pi \hbar }}} 
\end{equation}
and ${A_{\rho \rho }}({\bf{q}},E) =  - 2Im\chi _{\rho \rho }^R({\bf{q}},E)$. Then the self-energy Eq\eqref{selfenergyrho} can be rewritten using spectral function as 
\begin{equation}
\Sigma _N^{y}({\bf{k}},i{p_m}) =  - \frac{{{g^2}}}{{{\beta ^1}{L^6}}}\int\limits_{ - \infty }^{ + \infty } {\frac{{dE}}{{2\pi \hbar }}} \sum\limits_{{\bf{q}}n} {\frac{1}{{i{p_m} - i{\omega _n} - {E_{N{\bf{k}} - {\bf{q}}}}}}\frac{{{A_{\rho \rho }}({\bf{q}},E)}}{{i{\omega _n} - E}}} 
\label{selfenergyrho2}
\end{equation}
Carrying out the Matsubara frequency summation, we have
\begin{equation}
\frac{1}{\beta }\sum\limits_n {\frac{1}{{i{p_m} - i{\omega _n} - {E_{N{\bf{k}} - {\bf{q}}}}}}\frac{1}{{i{\omega _n} - E}}}  =  - \frac{{{n_B}(E) + 1 - {n_F}({E_{N{\bf{k}} - {\bf{q}}}})}}{{i{p_m} - {E_{N{\bf{k}} - {\bf{q}}}} - E}}
\end{equation}
We can simply by setting ${n_F}({E_{N{\bf{k}} - {\bf{q}}}}) = 0$ since neutrons are free particles without Fermi sea. Substitute back to Eq.\,\eqref{selfenergyrho2}, we have 
\begin{equation}
\Sigma _N^{y}({\bf{k}},i{p_m}) =  - \frac{{{g^2}}}{{{L^6}}}\int\limits_{ - \infty }^{ + \infty } {\frac{{dE}}{{\pi \hbar }}} \sum\limits_{\bf{q}} {\frac{{{n_B}(\omega ) + 1}}{{i{p_m} - {E_{N{\bf{k}} - {\bf{q}}}} - E}}} \chi ''({\bf{q}},E)
\end{equation}
After analytical continuation to real frequency $i{p_m} \to {E_0} + i\delta$, we have
\begin{equation}
\Sigma _N^{y}({\bf{k}},{E_0}) =  - \frac{{{g^2}}}{{{L^6}}}\int\limits_{ - \infty }^{ + \infty } {\frac{{dE}}{{\pi \hbar }}} \sum\limits_{\bf{q}} {\chi ''({\bf{q}},E)\frac{{{n_B}(E) + 1}}{{{E_0} - {E_{N{\bf{k}} - {\bf{q}}}} - E + i\delta }}} 
\end{equation}
To link back to the final form of dynamic structure factor and aim to be consistent to the classical derivation as much as possible, we define the dynamic structure factor, which can also be considered as the spectral function of the density-density correlation function in the following way: 
\begin{equation}
S({\bf{q}},E) \equiv \int\limits_{ - \infty }^{ + \infty } {I({\bf{q}},t)} {e^{iEt/\hbar }}dt = \int\limits_{ - \infty }^{ + \infty } {\left\langle {\rho ({\bf{q}},t)\rho ( - {\bf{q}},0)} \right\rangle } {e^{iEt/\hbar }}dt
\end{equation}
Then, we have 
\begin{equation}
    S({\bf{q}},E) =  - 2\left[ {{n_B}(E) + 1} \right]\chi ''({\bf{q}},E) =  + \left[ {{n_B}(E) + 1} \right]A({\bf{q}},E)
\end{equation}

Finally, we have the neutron self-energy in terms of the dynamic structure factor 
\begin{equation}
\Sigma _N^{y}({\bf{k}},{E_0}) =  + \frac{{{g^2}}}{{{L^6}}}\int\limits_{ - \infty }^{ + \infty } {\frac{{dE}}{{2\pi \hbar }}} \sum\limits_{\bf{q}} {\frac{{S({\bf{q}},E)}}{{{E_0} - {E_{N{\bf{k}} - {\bf{q}}}} - E + i\delta }}} 
\end{equation}
For neutron scattering, we only need the “on-shell” part since both incoming and outgoing neutron are both free neutron, i.e. ${E_0} = {E_{\rm{i}}} = {E_{N{\bf{k}}}} = \frac{{{\hbar ^2}{k^2}}}{{2{m_N}}}$, then, using the more conventional neutron scattering notations, calling incident and scattered final neutron momenta as $\bf{k_i}$ and $\bf{k_f}$, with related energy $E_i$ and $E_f$, respectively, and neutron momentum change ${\bf{Q}} = {{\bf{k}}_{\rm{i}}} - {{\bf{k}}_{\rm{f}}}$ Then, using the conservation of total number of states, we have density of state conservation
\begin{equation}
\int {{\rho _{{{\bf{k}}_{\rm{f}}}}}({E_{\rm{f}}})d{E_{\rm{f}}}}  = \sum\limits_{{{\bf{k}}_{\rm{f}}}} 1  = \frac{{{L^3}}}{{{{(2\pi )}^3}}}\int {{d^3}{{\bf{k}}_{\rm{f}}}}  = \frac{{{L^3}}}{{{{(2\pi )}^3}}}\int {k_{\rm{f}}^2d{k_{\rm{f}}}d\Omega }     
\end{equation}

$d\Omega$ is the solid angle around the final momentum $\bf{k_f}$. Since for non-relativistic neutrons we have ${E_{\rm{f}}} = \frac{{{\hbar ^2}k_{\rm{f}}^2}}{{2{m_N}}} \Rightarrow d{E_{\rm{f}}} = \frac{{{\hbar ^2}{k_{\rm{f}}}}}{{{m_N}}}d{k_{\rm{f}}}$.  Then using the monotonic behavior, we can eliminate the integration: 
\begin{equation}
{\rho _{{{\bf{k}}_{\rm{f}}}}}({E_{\rm{f}}}) = \frac{{{L^3}}}{{{{(2\pi )}^3}}}k_{\rm{f}}^2\frac{{d{k_{\rm{f}}}}}{{d{E_{\rm{f}}}}}d\Omega  = \frac{{{L^3}}}{{{{(2\pi )}^3}}}k_{\rm{f}}^2\frac{{{m_N}}}{{{\hbar ^2}{k_{\rm{f}}}}}d\Omega  = \frac{{{L^3}}}{{{{(2\pi )}^3}}}\frac{{{m_N}{k_{\rm{f}}}}}{{{\hbar ^2}}}d\Omega 
\end{equation}
after changing the integration variable, we have 
\begin{equation}
\begin{split}
\frac{\hbar }{{{\tau _{{{\bf{k}}_{\rm{i}}}}}}} &=  - 2{\mathop{\rm Im}\nolimits} \Sigma _N^{y}({{\bf{k}}_{\rm{i}}},{E_{\rm{i}}}) = \frac{{{g^2}}}{{{L^6}}}\sum\limits_{{{\bf{k}}_{\rm{f}}}} {S({\bf{Q}} = {{\bf{k}}_{\rm{i}}} - {{\bf{k}}_{\rm{f}}},E = {E_{\rm{i}}} - {E_{\rm{f}}})} \\&
 = \frac{{{g^2}}}{{{L^3}}}\int {\frac{{{m_N}{k_{\rm{f}}}}}{{8{\pi ^2}{\hbar ^2}}}d{E_{\rm{f}}}d{\Omega _{\rm{f}}}S({\bf{Q}} = {{\bf{k}}_{\rm{i}}} - {{\bf{k}}_{\rm{f}}},E = {E_{\rm{i}}} - {E_{\rm{f}}})} 
\end{split}
\end{equation}
The total cross section is normalized over incident particle flux $\Phi  = \frac{{\hbar {k_{\rm{i}}}}}{{{L^3}{m_N}}}$ and cross-section $\sigma  = \frac{{{1 \mathord{\left/
 {\vphantom {1 {{\tau _{\bf{k}}}}}} \right.
 \kern-\nulldelimiterspace} {{\tau _{\bf{k}}}}}}}{\Phi }$, finally we have the double differential cross section for nuclear scattering as 
\begin{equation}
\frac{{{d^2}{\sigma _{{\rm{nuc}}}}}}{{d{E_{\rm{f}}}d\Omega }} = \frac{{{k_{\rm{f}}}}}{{{k_{\rm{i}}}}}{\left( {\frac{{g{m_N}}}{{2\pi {\hbar ^2}}}} \right)^2}\frac{1}{{2\pi \hbar }}S({\bf{Q}},E)
\end{equation}
Comparing with conventional notation, for $g = \frac{{2\pi {\hbar ^2}}}{{{m_N}}}$ we have 
\begin{equation}
\frac{{{d^2}{\sigma _{{\rm{nuc}}}}}}{{d{E_{\rm{f}}}d\Omega }} = \frac{{{k_{\rm{f}}}}}{{{k_{\rm{i}}}}}\frac{1}{{2\pi \hbar }}S({\bf{Q}},E)
\end{equation}
This exactly reproduces the results in classical Fermi’s golden rule and concludes the derivation. 

The magnetic potential energy density for a neutron at position $\bf{r'}$ felt by the spin and orbital magnetism of an electron at position $\bf{r}$ can be written as 
\begin{equation}
V(\left. {{\bf{r'}}} \right|{\bf{r}}) =  - 2\gamma {\mu _B}{\mu _N}{{\bf{\sigma }}_N}({\bf{r'}}) \cdot \left( {{\nabla _{{\bf{r'}}}} \times \left[ {\frac{{\left( {{\bf{s}}({\bf{r}}) \times {\bf{\hat R}}} \right)}}{{{R^2}}}} \right] + \frac{{{{\bf{k}}_e}({\bf{r}}) \times {\bf{\hat R}}}}{{{R^2}}}} \right)
\end{equation}
where $\gamma  =  + 1.913$ is the dimensionless gyromagnetic ratio, $\mu_B$ and $\mu_N$ are the Bohr magneton and nuclear magneton, respectively, ${\bf{R}} \equiv {\bf{r'}} - {\bf{r}}$, and ${\bf{\hat R}} = {{\bf{R}} \mathord{\left/
 {\vphantom {{\bf{R}} {\left| {\bf{R}} \right|}}} \right.
 \kern-\nulldelimiterspace} {\left| {\bf{R}} \right|}}$, and neutron Pauli matrix (density) can be written as ${{\bf{\sigma }}_N}({\bf{r'}}) = \sum\limits_{\alpha \beta } {\Psi _{N\alpha }^ + ({\bf{r'}}){{\bf{\sigma }}_{N\alpha \beta }}{\Psi _{N\beta }}({\bf{r'}})}$, where $\alpha ,\beta  = 1,2$ are Pauli matrix indices, the (dimensionless) electron spin density operator ${\bf{s}}({\bf{r}})$ can be written as ${\bf{s}}({\bf{r}}) = \sum\limits_{\alpha \beta } {\psi _\alpha ^ + ({\bf{r}}){{\bf{s}}_{\alpha \beta }}{\psi _\beta }({\bf{r}})}$, electron orbital motion wavevector density is ${{\bf{k}}_e}({\bf{r}}) = \sum\limits_\alpha  {\psi _\alpha ^ + ({\bf{r}}){{\bf{k}}_e}{\psi _\alpha }({\bf{r}})} $ . Then, integrating over the space of the material, we have the total neutron-magnetic interaction Hamiltonian written as 
\begin{equation}
{H_I} =  - 2\gamma {\mu _B}{\mu _N}\int {{d^3}{\bf{r'}}{d^3}{\bf{r}}} \Psi _N^ + ({\bf{r'}}){{\bf{\sigma }}_N}{\Psi _N}({\bf{r'}}) \cdot \left( \begin{array}{l}
{\nabla _{{\bf{r'}}}} \times \left( {{\psi ^ + }({\bf{r}}){\bf{s}}\psi ({\bf{r}}) \times \frac{{\bf{R}}}{{{R^3}}}} \right)\\
 + {\psi ^ + }({\bf{r}}){{\bf{k}}_e}{I_{2 \times 2}}\psi ({\bf{r}}) \times \frac{{\bf{R}}}{{{R^3}}}
\end{array} \right)
\end{equation}
where both neutron field and electron field are written as two-component spinor forms, that ${\Psi _N}({\bf{r'}}) = {\left( {\begin{array}{*{20}{c}}
{{\Psi _{N \uparrow }}({\bf{r'}})}&{{\Psi _{N \downarrow }}({\bf{r'}})}
\end{array}} \right)^T},\psi ({\bf{r}}) = {\left( {\begin{array}{*{20}{c}}
{{\psi _ \uparrow }({\bf{r}})}&{{\psi _ \downarrow }({\bf{r}})}
\end{array}} \right)^T}$. Defining the Fourier transform of the electron field spinor ${c_{\bf{k}}} = \frac{1}{{{L^{3/2}}}}\int {{d^3}{\bf{r}}{e^{ - i{\bf{k}} \cdot {\bf{r}}}}\psi ({\bf{r}})}$ and 
projected magnetization operator ${{\bf{M}}_ \bot }({\bf{q}}) = \sum\limits_{{\bf{k'}}} {c_{{\bf{k'}}}^ + {{\bf{m}}_ \bot }({\bf{q}}){c_{{\bf{k'}} + {\bf{q}}}}}$ (${{\bf{m}}_ \bot }({\bf{q}}) = \frac{{{\bf{q}} \times {\bf{s}} \times {\bf{q}}}}{{{q^2}}} - i{{\bf{k}}_e}{I_{2 \times 2}} \times \frac{{\bf{q}}}{{{q^2}}}$), and using the fact that $4\gamma {\mu _B}{\mu _N} = \frac{{{\hbar ^2}\gamma {r_e}}}{{{m_p}}} \approx \frac{{{\hbar ^2}\gamma {r_e}}}{{{m_n}}}$ , the neutron magnetic interaction Hamiltonian can be written as 
\begin{equation}
{H_I} =  - \frac{{2\pi {\hbar ^2}\gamma {r_e}}}{{{m_N}{L^3}}}\sum\limits_{{\bf{qk}}} {\Psi _{N{\bf{k}} + {\bf{q}}}^ + {{\bf{\sigma }}_N}{\Psi _{N{\bf{k}}}} \cdot {{\bf{M}}_ \bot }({\bf{q}})} 
\label{intHmag}
\end{equation}
As carried before, with the interaction Eq.\,\eqref{intHmag}, the lowest order self-energy in Matsubara frequency domain can be written as 
\begin{equation}
\Sigma _{N\mu \nu }^{y}({\bf{k}},i{p_m}) =  - {\left( {\frac{{2\pi {\hbar ^2}\gamma {r_e}}}{{{m_N}{L^3}}}} \right)^2}\frac{1}{\beta }\sum\limits_{\alpha \beta }^{{\rm{spin}}} {\sum\limits_{i,j = 1}^3 {\sum\limits_{{\bf{q}}n} {\sigma _{N\mu \alpha }^iG_{N\alpha \beta }^{(0)}({\bf{k}} - {\bf{q}},i{p_m} - i{\omega _n})\sigma _{N\beta \nu }^j\chi _{MM \bot }^{ij}({\bf{q}},i{\omega _n})} } } 
\label{selfEM}
\end{equation}
where the $\Sigma _{N \uparrow  \uparrow }^{y}({\bf{k}},i{p_n})$ and $\Sigma _{N \downarrow  \downarrow }^{y}({\bf{k}},i{p_n})$ terms denotes the processes for the spin non-flip scattering, while the $\Sigma _{N \uparrow  \downarrow }^{y}({\bf{k}},i{p_n})$ and $\Sigma _{N \downarrow  \uparrow }^{y}({\bf{k}},i{p_n})$ are the spin flip scattering. The magnetic response function is defined as  
\begin{equation}
    \begin{split}
\chi _{MM \bot }^{ij}({\bf{k}},i{\omega _n}) &= \int\limits_0^\beta  {\chi _{MM \bot }^{ij}({\bf{k}},\tau ){e^{i{\omega _n}\tau }}d\tau } \\
\chi _{MM \bot }^{ij}({\bf{q}},{\tau _1} - {\tau _2}) &=  - {{\rm{T}}_\tau }{\left\langle {M_ \bot ^i({\bf{q}},{\tau _1})M_ \bot ^j( - {\bf{q}},{\tau _2})} \right\rangle _H} = \frac{1}{\beta }\sum\limits_n {\chi _{MM \bot }^{ij}({\bf{q}},i{\omega _n}){e^{ - i{\omega _n}({\tau _1} - {\tau _2})}}} 
    \end{split}
\end{equation}
The non-interacting neutron Green’s function is diagonal in spin space, i.e., $G_{N\alpha \beta }^{(0)}({\bf{k}} - {\bf{q}},i{p_m} - i{\omega _n}) = G_N^{(0)}({\bf{k}} - {\bf{q}},i{p_m} - i{\omega _n}){\delta _{\alpha \beta }}$, and then weight average the spin-up and spin-down components,
\begin{equation}
\sum\limits_{\alpha \beta }^{} {\sigma _{N\mu \alpha }^i{\delta _{\alpha \beta }}\sigma _{N\beta \nu }^j}  = \sum\limits_\alpha ^{} {\sigma _{N\mu \alpha }^i\sigma _{N\alpha \nu }^j}  = {\left( {\sigma _N^i\sigma _N^j} \right)_{\mu \nu }} = {\delta _{ij}}{I_{\mu \nu }} + i{\left( {{\varepsilon _{ijk}}{\sigma _k}} \right)_{\mu \nu }}
\end{equation}
For unpolarized neutrons, it is equal mixing of up-up $\uparrow\uparrow$ and down-down $\downarrow\downarrow$ components, and unpolarized  the averaged neutron self-energy $\Sigma _N^{y}({\bf{k}},i{p_n}) = \frac{1}{2}\left( {\Sigma _{N \uparrow  \uparrow }^{y}({\bf{k}},i{p_n}) + \Sigma _{N \downarrow  \downarrow }^{y}({\bf{k}},i{p_n})} \right)$. Then, 
\begin{equation}
\frac{1}{2}\left( {\sum\limits_{\alpha \beta }^{} {\sigma _{N \uparrow \alpha }^i{\delta _{\alpha \beta }}\sigma _{N\beta  \uparrow }^j}  + \sum\limits_{\alpha \beta }^{} {\sigma _{N \downarrow \alpha }^i{\delta _{\alpha \beta }}\sigma _{N\beta  \downarrow }^j} } \right) = {\delta _{ij}} + \frac{i}{2}{\rm{Tr}}\left( {{\varepsilon _{ijk}}{\sigma _k}} \right) = {\delta _{ij}}
\end{equation}
and the self-energy in Eq.\,\eqref{selfEM} for unpolarized neutron can be simplified as 
\begin{equation}
\Sigma _N^{y}({\bf{k}},i{p_m}) =  - {\left( {\frac{{2\pi {\hbar ^2}\gamma {r_e}}}{{{m_n}{L^3}}}} \right)^2}\frac{1}{\beta }\sum\limits_{{\bf{q}}n} {G_N^{(0)}({\bf{k}} - {\bf{q}},i{p_m} - i{\omega _n})\sum\limits_{j = 1}^3 {\chi _{MM \bot }^{jj}({\bf{q}},i{\omega _n})} } 
\end{equation}
Now following the same procedure of many-body derivation on nuclear scattering, we define the projected spin spectral function $A_{MM \bot }^j({\bf{q}},E)$ as 
\begin{equation}
\chi _{MM \bot }^{jj}({\bf{q}},i{\omega _r}) = \int\limits_{ - \infty }^{ + \infty } {\frac{{A_{MM \bot }^j({\bf{q}},E)}}{{i{\omega _r} - E}}\frac{{dE}}{{2\pi \hbar }}} 
\end{equation}
and further define the projected magnetic correlation function (magnetic dynamic structure factor)  
\begin{equation}
    \begin{split}
S_{MM \bot }^j\left( {{\bf{q}},E} \right) &= \mathop \smallint \limits_{ - \infty }^\infty  \left\langle {M_ \bot ^j\left( {{\bf{q}},t} \right)M_ \bot ^j\left( { - {\bf{q}},0} \right)} \right\rangle {e^{iEt/\hbar }}dt\\&
 = \left[ {{n_B}(E) + 1} \right]A_{MM \bot }^j({\bf{q}},E) =  - 2\left[ {{n_B}(E) + 1} \right]{\mathop{\rm Im}\nolimits} \chi _{MM \bot }^{jj,R}({\bf{q}},E)
    \end{split}
\end{equation}
Then substituting them back to the self-energy for unpolarized neutron, we have 
\begin{equation}
\Sigma _N^{y}({\bf{k}},i{p_m}) =  + {\left( {\frac{{2\pi {\hbar ^2}\gamma {r_e}}}{{{m_n}{L^3}}}} \right)^2}\int\limits_{ - \infty }^{ + \infty } {\frac{{dE}}{{2\pi \hbar }}} \sum\limits_{j = 1}^3 {\sum\limits_{\bf{q}} {\frac{{S_{MM \bot }^j\left( {{\bf{q}},E} \right)}}{{i{p_m} - E - {E_{N{\bf{k}} - {\bf{q}}}}}}} }
\end{equation}
where the magnetic double differential cross section 
\begin{equation}
\frac{{{d^2}{\sigma _{{\rm{mag}}}}}}{{d{E_{\rm{f}}}d\Omega }} = \frac{{{k_{\rm{f}}}}}{{{k_{\rm{i}}}}}{\left( {\gamma {r_e}} \right)^2}\frac{1}{{2\pi \hbar }}\sum\limits_{j = 1}^3 {S_{MM \bot }^j({\bf{Q}},E)} 
\end{equation}
An alternative expression for cross section is to define a full vector magnetization operator $\bf{M(q)}$  as
\begin{equation}
    \begin{split}
{\bf{M}}({\bf{Q}}) &= \sum\limits_{\bf{k}} {c_{\bf{k}}^ + {\bf{m}}({\bf{Q}}){c_{{\bf{k}} + {\bf{Q}}}}}\\{\bf{M}}({\bf{Q}}) &= \sum\limits_l {{e^{i{\bf{Q}} \cdot {{\bf{R}}_l}}}\left( {{{\bf{s}}_l} + \frac{i}{{{Q^2}}}\left( {{{\bf{k}}_{e,l}} \times {\bf{Q}}} \right)} \right)}
    \end{split}
\end{equation}
for itinerant and localized magnetic moments, respectively, and ${\bf{m}}({\bf{Q}}) = {\bf{s}} - i{{\bf{k}}_e} \times \frac{{\bf{Q}}}{{{Q^2}}}{I_{2 \times 2}}$satisfies the following identity ${\bf{Q}} \times {\bf{m}}({\bf{Q}}) \times {\bf{Q}} = {Q^2}{\bf{m}}({\bf{Q}}) - \left( {{\bf{Q}} \cdot {\bf{m}}({\bf{Q}})} \right){\bf{Q}} = {Q^2}{{\bf{m}}_ \bot }({\bf{Q}})$. Then, projected magnetization operator ${{\bf{M}}_ \bot }({\bf{Q}})$ and full magnetization operator ${\bf{M}}({\bf{Q}})$ are related as 
\begin{equation}
M_ \bot ^i({\bf{Q}}) = \sum\limits_j {\left( {{\delta _{ij}} - {{\hat Q}_i}{{\hat Q}_j}} \right)} {M^j}({\bf{Q}})
\end{equation}
which is valid. As a result, if we define a magnetic dynamical structure factor tensor 
\begin{equation}
S_{MM}^{jp}({\bf{Q}},E) = \int\limits_{ - \infty }^{ + \infty } {dt{e^{ + iEt/\hbar }}} \left\langle {{M^j}({\bf{Q}},t){M^p}( - {\bf{Q}},0)} \right\rangle 
\end{equation}
Combining these equations, the double differential cross section finally can be written as 
\begin{equation}
\frac{{{d^2}{\sigma _{{\rm{mag}}}}}}{{d\Omega dE}} = \frac{{{k_f}}}{{{k_i}}}{\left( {\gamma {r_e}} \right)^2}\frac{1}{{2\pi \hbar }}\sum\limits_{jp = 1}^3 {\left( {{\delta _{jp}} - {{\hat Q}_j}{{\hat Q}_p}} \right)S_{MM}^{jp}({\bf{Q}},E)}
\end{equation}

\section{C. Anomalous dynamical structure factor from electron-phonon interaction}
In this section, we show the general workflow to compute the anomalous dynamical structure factor. The full material Hamiltonian H can be written as 
\begin{equation}
H = {H_0} + {H_{{\rm{int}}}} 
\end{equation}
rewriting it as “interaction picture”, where the interaction picture here means the separation of “non-interacting” part of materials Hamiltonian ($H_0$) and the interaction part ($H_{int}$). Defining the interaction picture time imaginary time evolution $U(\tau ) = {T_{\tau '}}\exp \left( { - \int_0^\tau  {d\tau '{{\hat H}_{{\rm{int}}}}(\tau ')} } \right)$, the anomalous response function in imaginary time can be written as 
\begin{equation}
    \begin{split}
\chi _{\rho M}^j({\bf{q}},{\tau _1} - {\tau _2}) =  - {{\rm{T}}_\tau }{\left\langle {{\rho _{\bf{q}}}({\tau _1})M_{ - {\bf{q}} \bot }^j({\tau _2})} \right\rangle _H}\\
 =  - \frac{{{{\rm{T}}_\tau }{{\left\langle {U(\beta ){{\hat \rho }_{\bf{q}}}({\tau _1})\hat M_{ - {\bf{q}} \bot }^j({\tau _2})} \right\rangle }_{{H_0}}}}}{{{{\left\langle {U(\beta )} \right\rangle }_{{H_0}}}}}
    \end{split}
    \label{eq:generalADS}
\end{equation}
where all the “hat” operators are in interaction picture w.r.t. non-interacting part of materials Hamiltonian, i.e. $\hat A(\tau ) = {e^{ + \tau {H_0}}}A{e^{ - \tau {H_0}}}$. Here We limit our discussion first for the system only with the weak electron-phonon interaction to concentrate on probing the electronic properties, we can perturbatively expand the $U(\beta )$, $U(\beta ) \approx 1 - \int_0^\beta  {d\tau } {\hat H_{{\rm{e - ph}}}}(\tau )$ , while for nuclear density, we have  ${\rho _{\bf{q}}} \approx \sum\limits_l {{b_l}{e^{ - i{\bf{q}} \cdot {\bf{R}}_l^0}}} \left( {1 - i{\bf{q}} \cdot {\bf{u}}({\bf{R}}_l^0)} \right)$
\begin{equation}
{\hat \rho _{\bf{q}}}({\tau _1}) = \sum\nolimits_{l = 1}^N {{b_l}{e^{ - i{\bf{q}} \cdot {{{\bf{\hat R}}}_l}({\tau _1})}}}  \approx \sum\limits_l {{b_l}{e^{ - i{\bf{q}} \cdot {\bf{R}}_l^0}}} \left( {1 - i{\bf{q}} \cdot {{{\bf{\hat u}}}_l}({\tau _1})} \right)
\end{equation}
Substituting back to the anomalous structure factor, we have 
\begin{equation}
\chi _{\rho M}^j({\bf{q}},{\tau _1} - {\tau _2}) \approx  - i{{\rm{T}}_\tau }\int_0^\beta  {d\tau \sum\nolimits_{l = 1}^N {{b_l}{e^{ - i{\bf{q}} \cdot {\bf{R}}_l^0}}{{\left\langle {{{\hat H}_{{\rm{e - ph}}}}(\tau ){\bf{q}} \cdot {{{\bf{\hat u}}}_l}({\tau _1})\hat M_{ - {\bf{q}} \bot }^j({\tau _2})} \right\rangle }_{{H_0}}}} } 
\end{equation}
where we only keep to the first order  ${\hat H_{{\rm{e - ph}}}}(\tau )$ in  $U(\beta )$, where ${\left\langle {U(\beta )} \right\rangle _{{H_0}}} \approx 1$. 
Performing explicit calculation, assuming a single mode phonon and average nuclear scattering length density $\Bar{b}$,
\begin{equation}
\begin{split}
\chi _{\rho M}^j({\bf{q}},{\tau _1} - {\tau _2}) &\approx {{\rm{T}}_\tau }\int_0^\beta  {d\tau \sum\nolimits_{l = 1}^N {{b_l}{e^{ - i{\bf{q}} \cdot {\bf{R}}_l^0}}{{\left\langle {{{\hat H}_{{\rm{e - ph}}}}(\tau )( - i){\bf{q}} \cdot {{{\bf{\hat u}}}_l}({\tau _1})\hat M_{ - {\bf{q}} \bot }^j({\tau _2})} \right\rangle }_{{H_0}}}} } \\&
 = \frac{{{N^{1/2}}}}{V}\int_0^\beta  {d\tau \sum\nolimits_{l = 1}^N {{b_l}{e^{ - i{\bf{q}} \cdot {\bf{R}}_l^0}}{{\rm{T}}_\tau }{{\left\langle \begin{array}{l}
\sum\limits_{{\bf{k'q'}}\sigma } {{g_{{\bf{q'}}}}c_{{\bf{k'}} + {\bf{q'}}\sigma }^ + (\tau ){c_{{\bf{k'}}\sigma }}(\tau )(a_{{\bf{q'}}}^{}(\tau ) + a_{ - {\bf{q'}}}^ + (\tau ))} \\
 \times \sum\limits_{{\bf{q''}}} {\sqrt {\frac{\hbar }{{2MN{\omega _{{\bf{q''}}}}}}} \left( {a_{{\bf{q''}}}^{}({\tau _1}) + a_{ - {\bf{q''}}}^ + ({\tau _1})} \right){e^{i{\bf{q''}} \cdot {\bf{R}}_l^0}}\left( { - i{\bf{q}} \cdot {\varepsilon _{{\bf{q''}}}}} \right)} \\
 \times \sum\limits_{{\bf{k''}}\sigma '\sigma ''} {c_{{\bf{k''}}\sigma '}^ + ({\tau _2}){\bf{m}}_{ \bot \sigma '\sigma ''}^j( - {\bf{q}}){c_{{\bf{k''}} - {\bf{q}}\sigma ''}}({\tau _2})} 
\end{array} \right\rangle }_{{H_0}}}} }
\end{split}
\end{equation}
If we use that $\sum\nolimits_{l = 1}^N {{b_l}{e^{ - i{\bf{q}} \cdot {\bf{R}}_l^0}}{e^{i{\bf{q''}} \cdot {\bf{R}}_l^0}}}  = N\bar b{\delta _{{\bf{qq''}}}}$, $\bar b$ is the average nuclear scattering length density and contract the phonon Green’s function, we have
\begin{equation}
    \begin{split}
\chi _{\rho M}^j({\bf{q}},{\tau _1} - {\tau _2}) = \frac{N}{V}\bar b\int_0^\beta  {d\tau \sum\limits_{\scriptstyle{\bf{k'}}\sigma \hfill\atop
\scriptstyle{\bf{k''}}\sigma '\sigma ''\hfill} {{g_{ - {\bf{q}}}}\sqrt {\frac{\hbar }{{2M{\omega _{\bf{q}}}}}} \left( { - i{\bf{q}} \cdot {\varepsilon _{\bf{q}}}} \right){\bf{m}}_{ \bot \sigma '\sigma ''}^j( - {\bf{q}})} } \\
 \times {{\rm{T}}_\tau }{\left\langle {c_{{\bf{k'}} - {\bf{q}}\sigma }^ + (\tau ){c_{{\bf{k'}}\sigma }}(\tau )c_{{\bf{k''}}\sigma '}^ + ({\tau _2}){c_{{\bf{k''}} - {\bf{q}}\sigma ''}}({\tau _2})} \right\rangle _0}\left( { - 1} \right)D_{ - {\bf{q}}}^{(0)}(\tau  - {\tau _1})
    \end{split}
    \label{ANUBISGeneral}
\end{equation}
By far, Eq.\,\eqref{ANUBISGeneral} is the most general formula for electron-phonon interaction, with no assumptions on the types and directions of magnetism and q vector being made, and allows spin flip in electron propagators. 

Similar to the magnetic scattering discussed in the previous section, Both the spin and 
the orbital motion contributes to the signal through ${\bf{m}}_{ \bot \sigma '\sigma ''}^j( - {\bf{q}})$. We have mostly discussed the spin contribution in the main text while the orbital contribution will be briefly discussed here. For the contribution of the orbital motion of the electron, as we mentioned neutrons are seldom used in detecting this even in conventional settings, we have ${\bf{m}}_{ \bot ,{\rm{orb}} \uparrow  \uparrow }^z( - {\bf{q}}) =  + {\bf{m}}_{ \bot ,{\rm{orb}} \downarrow  \downarrow }^z( - {\bf{q}})$ with assumed constant $\bf{k_e}$, therefore 
\begin{equation}
    \begin{split}
\chi _{\rho M}^z({\bf{q}},i{\omega _n}) = \frac{N}{V}\bar b{g_{ - {\bf{q}}}}\sqrt {\frac{\hbar }{{2M{\omega _{\bf{q}}}}}} \left( { - i{\bf{q}} \cdot {\varepsilon _{\bf{q}}}} \right)\\
 \times {\bf{m}}_{ \bot ,{\rm{orb}} \uparrow  \uparrow }^z({\bf{q}})\left( {\chi _{{\bf{q}} \uparrow }^{ee}(i{\omega _n}) + \chi _{{\bf{q}} \downarrow }^{ee}(i{\omega _n})} \right)D_{ - {\bf{q}}}^{(0)}( - i{\omega _n})
    \end{split}
\end{equation}
The contribution usually is small and considered negligible so we leave this to be explored further. 

\section{D. ANUBIS signal Analysis}
\subsection{Normal Metal}
Starting from Eq.\,\eqref{eq:ano-MResponse}, simplifying the polarization part for the Fermi electron gas with the help of weak B field approximation and the low-temperature approximation for the Fermi statistics,  
\begin{equation}
    \begin{split}
\chi _{{\bf{q}} \uparrow }^{ee}(i{\omega _n}) - \chi _{{\bf{q}} \downarrow }^{ee}(i{\omega _n}) = \sum\limits_{\bf{k}} {\frac{{{n_F}({\varepsilon _{{\bf{k}} \uparrow }}) - {n_F}({\varepsilon _{{\bf{k}} - {\bf{q}} \uparrow }})}}{{ - i{\omega _n} + {\varepsilon _{{\bf{k}} \uparrow }} - {\varepsilon _{{\bf{k}} - {\bf{q}} \uparrow }}}}}  - \sum\limits_{\bf{k}} {\frac{{{n_F}({\varepsilon _{{\bf{k}} \downarrow }}) - {n_F}({\varepsilon _{{\bf{k}} - {\bf{q}} \downarrow }})}}{{ - i{\omega _n} + {\varepsilon _{{\bf{k}} \downarrow }} - {\varepsilon _{{\bf{k}} - {\bf{q}} \downarrow }}}}} \\
 = \sum\limits_{\bf{k}} {\frac{{\left( { - 2{\mu _B}{B_z}} \right)\left( {\frac{{d{n_F}({\varepsilon _{\bf{k}}})}}{{d{\varepsilon _{\bf{k}}}}} - \frac{{d{n_F}({\varepsilon _{{\bf{k}} - {\bf{q}}}})}}{{d{\varepsilon _{{\bf{k}} - {\bf{q}}}}}}} \right)}}{{ - i{\omega _n} + {\varepsilon _{\bf{k}}} - {\varepsilon _{{\bf{k}} - {\bf{q}}}}}}}  = \sum\limits_{\bf{k}} {\frac{{\delta ({\varepsilon _{\bf{k}}}) - \delta ({\varepsilon _{{\bf{k}} - {\bf{q}}}})}}{{ - i{\omega _n} + {\varepsilon _{\bf{k}}} - {\varepsilon _{{\bf{k}} - {\bf{q}}}}}}} 2{\mu _B}{B_z} \equiv 2{\mu _B}{B_z} C_{M, \rho}  
    \end{split}
    \label{eq:Pexpansion}
\end{equation}
Here we have used 
${\varepsilon _{{\bf{k}} \uparrow }} - {\varepsilon _{{\bf{k}} - {\bf{q}} \uparrow }} = 
{\varepsilon _{{\bf{k}} \downarrow }} - {\varepsilon _{{\bf{k}} - {\bf{q}} \downarrow }}$. Redefine $\tilde{q} =  q / k_F$ and $\nu= \omega / {\hbar k_F v_F} $ and calculate the summation through the integral over the whole k-space. In the finite frequency/momentum case, the dielectric response can be written in terms of the Lindhard function as
\begin{equation}
\begin{split}
     \Re C_{M, \rho}(\tilde{q}, \nu) =& \frac{1}{\tilde{q}} \ln \left \vert \frac{\nu - 2\tilde{q} + \tilde{q}^2}{\nu + 2\tilde{q} + \tilde{q}^2} \right \vert + \left ( (\tilde{q}, \nu ) \mapsto (-\tilde{q}, -\nu )\right ) \\
     \Im C_{M, \rho}(\tilde{q}, \nu) =& \frac{- i \pi}{\tilde{q}} \Theta (2\tilde{q} - |\nu + q^2|) - (\nu \mapsto - \nu)
\end{split}
\end{equation}

\subsection{Topological Semimetal}
For topological semimetal under the magnetic field, we have Hamiltonian 
\begin{align}
    \mathcal{H} = \hbar v_F \vb{k} \cdot \bm{\tau} + \mu_B B_z \sigma_z
\end{align}
where $\tau$ is pseudo spin, we can block diagonalize the matrix and calculate the green function as 
\begin{align}
    G = \begin{pmatrix}
        G_\uparrow& 0 \\ 0 & G_\downarrow
    \end{pmatrix}
\end{align}
where $\epsilon_{\pm, \tau}(k) = \epsilon_\tau(k) \pm \mu_B B$ and $\epsilon_\tau = \tau \hbar v_F |\vb{k}|$ $\tau \in \{+1, -1\}$. To analyze the ANUBIS signal of Eq.\,\eqref{eq:ano-MResponse} for semimetal, we do a similar expansion,

\begin{equation}
    \begin{split}
C_{M, \rho} = \chi _{{\bf{q}} \uparrow }^{ee}(i{\omega _n}) - \chi _{{\bf{q}} \downarrow }^{ee}(i{\omega _n}) &= \sum_{\alpha, \beta \in \pm 1}\left[\sum\limits_{\bf{k}}F_{\alpha,\beta}\left(\bf{k},\bf{k-q}\right) {\frac{{{n_F}({\varepsilon _{{\bf{k}} \uparrow }}) - {n_F}({\varepsilon _{{\bf{k}} - {\bf{q}} \uparrow }})}}{{ - i{\omega _n} + {\varepsilon _{{\bf{k}} \uparrow }} - {\varepsilon _{{\bf{k}} - {\bf{q}} \uparrow }}}}}  - \sum\limits_{\bf{k}}F_{\alpha,\beta}\left(\bf{k},\bf{k-q}\right) {\frac{{{n_F}({\varepsilon _{{\bf{k}} \downarrow }}) - {n_F}({\varepsilon _{{\bf{k}} - {\bf{q}} \downarrow }})}}{{ - i{\omega _n} + {\varepsilon _{{\bf{k}} \downarrow }} - {\varepsilon _{{\bf{k}} - {\bf{q}} \downarrow }}}}}\right] \\& = \sum_{\alpha, \beta \in \pm 1}\sum\limits_{\bf{k}}F_{\alpha,\beta}\left(\bf{k},\bf{k-q}\right) {\frac{{\delta ({\varepsilon _{\bf{k}}}) - \delta ({\varepsilon _{{\bf{k}} - {\bf{q}}}})}}{{ - i{\omega _n} + {\varepsilon _{\bf{k}}} - {\varepsilon _{{\bf{k}} - {\bf{q}}}}}}} 2{\mu _B}{B_z} \equiv 2{\mu _B}{B_z}\sum_{\alpha, \beta \in \pm 1}C_{M, \rho}^{\alpha,\beta}
\end{split}
\end{equation}
Here $\alpha$ and $\beta$ denote the band indices, the quantity $F_{\alpha,\beta}\left(\bf{k},\bf{k-q}\right) = \frac{1}{2}\left(1 + \alpha \beta \frac{\bm{k}\cdot(\bm{k} - \bm{q})}{|\bm{k}||\bm{k  q}|}\right)$ comes from the overlap of eigenstates, which has the same expression as that in 2D graphene \cite{zhou2015plasmon,ando2006screening}.

The dielectric response can be calculated in a similar manner and yield
\begin{equation}
\begin{split}
     \Re C^{\alpha\beta}_{M, \rho}(\tilde{q}, \nu) = &  \frac{\Theta(\alpha \mu)}{2} + \frac{1}{2} \frac{ \Theta(\alpha \mu)}{\tilde{q}} \left( \frac{(\nu - \alpha)^2}{2} - \alpha (\nu - \alpha) + \frac{1-\tilde{q}^2}{2} \right)  \\ &\times \ln \left \vert \frac{\nu - \alpha + \beta |1+\tilde{q}|}{\nu - \alpha + \beta |1-\tilde{q}|}\right \vert   \\&+ \left ( (\nu, \tilde{q}, \alpha, \beta) \mapsto  (-\nu, -\tilde{q}, \beta, \alpha)\right)   \\
     \Im C^{\alpha\beta}_{M, \rho}(\tilde{q}, \nu) = & \frac{i \pi}{2 \tilde{q}}  \Theta(\alpha \mu) \beta \left (\alpha ( \nu - \alpha) - \frac{1}{2} (\nu - \alpha)^2 - \frac{1 - \tilde{q}^2}{2} \right ) \\ & \times\mathbbm{1}_{\frac{\alpha - \nu}{\beta} \in [|1-\tilde{q}|, |1+\tilde{q}|]} \\&+ \left ( (\nu, \tilde{q}, \alpha, \beta) \mapsto  (-\nu, -\tilde{q}, \beta, \alpha)\right) 
     \end{split}
\end{equation}

\end{document}